\documentclass[prb,twocolumn]{revtex4-1}

\usepackage{amsfonts,amsmath,amsthm,amssymb}
\usepackage{mathrsfs}
\usepackage[mathscr]{eucal}
\usepackage{xcolor}
\usepackage{graphicx}
\usepackage{hyperref}
\definecolor{linkcolour}{rgb}{0,0.2,0.6}
\hypersetup{colorlinks,breaklinks,
		linkcolor=linkcolour,citecolor=linkcolour,
        filecolor=linkcolour, urlcolor=linkcolour}

\usepackage{bm}
\usepackage{tikz}
\usepackage{bbm}

\newcommand\ham{\mathcal{H}}
\newcommand\ver{\mathcal{A}_\beta}
\newcommand\lr[1]{\left(#1\right)}
\newcommand\lrs[1]{\left[#1\right]}
\newcommand\cint[1]{\frac{\de #1}{2\pi i}}
\newcommand\vd{{\vphantom{\dagger}}}

\newcommand\blam{{\bm{\lambda}}}
\newcommand\bmu{{\bm{\mu}}}
\newcommand\bgam{{\bm{\Gamma}}}
\newcommand\bxi{{\bm{\Xi}}}
\newcommand\order[1]{\mathcal{O}\left(#1\right)}
\newcommand\de{\textrm{d}}
\newcommand\De{\textrm{D}}
\newcommand\eq[1]{\begin{equation}#1\end{equation}}
\newcommand\alg[1]{\begin{align}#1\end{align}}
\newcommand\tr[1]{\textrm{#1}}
\newcommand\kket[1]{\bm{\left|}#1\right\}}
\newcommand\ket[1]{\left|#1\right\rangle}
\newcommand\bbra[1]{\left\{#1\bm{\right|}}
\newcommand\bra[1]{\left\langle#1\right|}
\newcommand\bbrakket[2]{\left\{#1\middle|#2\right\}}

\begin{document}

\title{Effective Edge State Dynamics in the Fractional Quantum Hall Effect}
\author{R.~Fern}
\affiliation{Rudolf Peierls Centre for Theoretical Physics, University of Oxford, OX1 3PU}
\author{R.~Bondesan}
\affiliation{Rudolf Peierls Centre for Theoretical Physics, University of Oxford, OX1 3PU}
\author{S.~H.~Simon}
\affiliation{Rudolf Peierls Centre for Theoretical Physics, University of Oxford, OX1 3PU}

\begin{abstract}
We consider the behaviour of quantum Hall edges away from the Luttinger liquid fixed point that occurs in the low energy, large system limit.
Using the close links between quantum Hall wavefunctions and conformal field theories we construct effective Hamiltonians from general principles and then constrain their forms by considering the effect of bulk symmetries on the properties of the edge.
In examining the effect of bulk interactions on this edge we find remarkable simplifications to these effective theories which allow for a very accurate description of the low-energy physics of quantum Hall edges relatively far away from the Luttinger liquid fixed point, and which apply to small systems and higher energies.
\end{abstract}

\maketitle

\section{Introduction}

The edges of quantum Hall systems are remarkable.
This boundary is the paradigmatic example of a chiral quantum liquid\cite{chang1996observation, wen1990chiral}, a one-dimensional system in which the transport is only in one direction\cite{chang2003chiral, kane1992transport, giamarchi2004quantum} (the chirality being due to the breaking of time reversal symmetry in the presence of a magnetic field).
Furthermore, given that the bulk of quantum Hall systems are gapped, these edge modes constitute the only low-energy degrees of freedom in these systems, and therefore mediate the fascinating transport properties for which the quantum Hall effect is so well known\cite{halperin1982quantized, girvin1999quantum}.

The dynamics of these modes has been a topic of great interest since the classic papers of Wen\cite{wen1990chiral, wen1992theory, wen1995topological, wen1991gapless, wen1990electrodynamical}.
In these works it was found that the low-energy dynamics of edge modes in the thermodynamic limit corresponds exactly to a chiral linear Luttinger liquid.
Linearity means that the modes have some dispersion $\omega(k)$ which is linear in the wavenumber, $k$, i.e, $\omega=vk$, and chirality implies that $k>0$.

For generic systems this linear picture is only true in the scaling limit of low energies and large system sizes.
However, in the presence of a set of special, model interactions, and when the confinement is finely-tuned, the dispersion of the edge modes is linear regardless of system size all the way to high energies\cite{haldane1985finite, milovanovic1996edge, read2009conformal}.
Away from these special cases one must consider the effect of irrelevant contributions, which introduce nonlinearities such as nontrivial dispersion or scattering processes between modes.

In this work we will consider exactly this non-ideal case, which can be characterised by anharmonic confinement or interactions and will, in general, have a far richer, nonlinear edge structure.
This has been discussed in numerous works such as Refs.~\onlinecite{cappelliw1+, price2014fine, price2017nonlinear, price2015quantum, cooper2015signatures, fern2017quantum,bettelheim2006nonlinear, wiegmann2012nonlinear}.
To analyse these nonlinear effects we will take the ideal case where the edge dispersion is precisely linear, with given Hamiltonian $\ham_\tr{Parent}$, and perturb it with $\delta\ham$, thus moving it towards something more realistic.
We then construct an effective field theory for the perturbed system.
In doing so we generate a mapping from a perturbation acting upon the whole bulk of the droplet, $\delta\ham$, onto a low-energy effective Hamiltonian which resides only on the edge.

In order to constrain the field theory, we conjecture it to be local and impose upon it the symmetries of the perturbation using a construction inspired by work from J.~Dubail, N.~Read and E.~Rezayi\cite{dubail2012edge} and revisited recently in Ref.~\onlinecite{fern2018structure}.
We find that this procedure is especially fruitful as it maps symmetries of the perturbations, such as rotational and translational invariance, to powerful constraints on the effective Hamiltonian’s form.
We illustrate this mapping from bulk interactions to their effect on the edge for the Laughlin\cite{laughlin1983anomalous} and Moore-Read\cite{moore1991nonabelions} quantum Hall states, though the procedure could in principle be generalised to more exotic quantum Hall states, such as the Read-Rezayi states\cite{read1999beyond} or any other state which can be expressed by a conformal field theory\cite{francesco2012conformal, hansson2017quantum}.

We will see that this effective description of the edge dynamics is accurate for short-range interactions and confinements close to quadratic.
These two conditions make our work particularly applicable to potential cold atom realisations of the quantum Hall effect where the interactions between atoms are generally short-range, perhaps even hard-core, the confinement can be readily tuned to a simple quadratic, and the number of particles can be quite small\cite{cooper2013reaching, sorensen2005fractional, regnault2003quantum}.
In this regime our effective theories prove to be extremely good at capturing the effects of finite size and non-ideal interactions on the edge behaviour.

We begin in section \ref{theory} with a recap of quantum Hall edges and their construction in terms of conformal blocks.
We then introduce the concept of an effective Hamiltonian in section \ref{hamiltonians}, describe how this can be expressed as a field theory on the edge of our system and discuss the effect of symmetries.
We then use these powerful results in section \ref{results} for the Laughlin and Moore-Read wavefunctions to propose generic theories for the edge dynamics induced by non-ideal bulk interactions.
Finally, we present numerics in support of these claims in section \ref{numerics}, showing the excellent agreement between finite-size exact diagonalisation and our effective edge theories.

\section{Theoretical Background}
\label{theory}

In order to make progress we will use the close links between quantum Hall wavefunctions and conformal field theories.
As such, we begin this section by introducing the concepts of parent Hamiltonians for the Laughlin state, which allows us to make a precise statement of our problem.
We will then discuss the formation of the edge state wavefunctions for the Laughlin state in terms of conformal blocks, a construction which we shall make judicial use of going forward.

\subsection{Quantum Hall Edges}

\subsubsection{Trial Wavefunctions}

A quantum Hall system is one in which charged particles are confined to two dimensions in the presence of a perpendicular magnetic field\cite{girvin1999quantum}.
The Hamiltonian of such a system containing $N$ particles is
	\alg{\ham = & \sum_{i=1}^N\frac{(\bm{p}_i-q\bm{A})^2}{2m_q}
		 + \sum_{1\le i<j\le N}V(|\bm{r}_i-\bm{r}_j|)
		 + \sum_{i=1}^NU(|\bm{r}_i|)}
where $\bm{r}_i$ and $\bm{p}_i$ are the positions and momenta of charge-$q$ particles whose effective mass is $m_q$.
These particles are acted on by a magnetic field, given by the vector potential $\bm{A}$, they interact via some generic two-body interactions, $V(|\bm{r}_i-\bm{r}_j|)$, and are placed in some confining trap $U(|\bm{r}_i|)$.
Note that in this work we will only consider rotationally symmetric Hamiltonians on the plane.

When the interactions are trivial this Hamiltonian gives the integer quantum Hall effect, which is characterised by a series of $\textit{Landau levels}$ separated by a constant gap of $\hbar\omega_c$ where $\omega_c$ is the cyclotron frequency.
The introduction of interactions introduces new gaps into this spectrum of order $V$, thus generating a hierarchy of fractional quantum Hall states.
In both cases, the total filling of these levels is determined by the density of particles, which given a fixed number, $N$, is determined by the confinement of the particles, $U$.
We will work in the limit
	\eq{\hbar\omega_c \gg V \gg U.}

So long as the number of electrons in the system fills an integer number of Landau levels, a single-particle assumption remains reasonably accurate\cite{von1986quantized, tsui1982two}.
When the filling, $\nu$, is fractional however, the problem is far more difficult.
Nevertheless, a variety of extremely well-educated guesses have been proposed over the years which approximate the true ground states of these fractionally-filled systems remarkably well.
The first of these was the Laughlin wavefunction\cite{laughlin1983anomalous},
	\eq{\Psi(\bm{z}) = \prod_{1\le i<j\le N}\lr{z_i-z_j}^\beta\exp\lr{-\sum_{i=1}^N\frac{|z_i|^2}{4\ell_B^2}}
			\label{Laughlin state}}
where $z_i\in\mathbbm{C}$ are the particle positions and $\ell_B$ the magnetic length.
This describes the fractional quantum Hall effect at fillings $\nu=1/\beta$ for non-negative integer $\beta$.
Subsequent proposals include the Moore-Read state, the Read-Rezayi series, the Haffnian and the Gaffnian\cite{moore1991nonabelions, read1999beyond, green2002strongly, simon2007construction}.
In each case the wavefunction is a holomorphic polynomial (up to Gaussian factors) of the positions where the power attached to any $z_i$ is the angular momentum of that $i^\tr{th}$ particle.

\subsubsection{Edge States}

We are interested in the low-energy states on top of these trial wavefunctions, which are formed by adding angular momentum to the ground state without changing the (bosonic or fermionic) symmetry of the wavefunction or increasing its energy too much.
The specific construction in terms of holomorphic polynomials will be discussed in section \ref{effective descriptions} but there exists an intuitive picture for the Laughlin state thanks to Wen\cite{wen1992theory}.

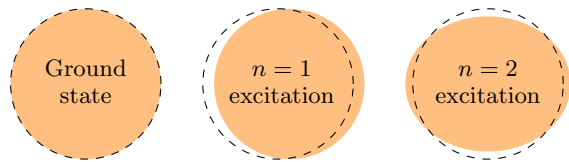
\begin{figure}[ht!]
	\centering
	\begin{tikzpicture}
		\fill [orange,opacity=0.5] (0,0) circle(1);
		\draw [dashed] (0,0) circle (1);
		\node at (0,0) {\begin{tabular}{c}Ground\\state\end{tabular}};
	\end{tikzpicture}
	$\quad$
	\begin{tikzpicture}
		\fill [orange,opacity=0.5] (0.15,0) circle (1);
		\draw [dashed] (0,0) circle (1);
		\node at (0.05,0) {\begin{tabular}{c}$n=1$\\excitation\end{tabular}};
	\end{tikzpicture}
	$\quad$
	\begin{tikzpicture}
		\fill [orange,opacity=0.5] (0,0) ellipse (1.1cm and 0.9cm);
		\draw [dashed] (0,0) circle (1);
		\node at (0,0) {\begin{tabular}{c}$n=2$\\excitation\end{tabular}};
	\end{tikzpicture}
	\caption{A cartoon picture of edge modes of Laughlin droplets.
		These low-energy excitations can be visualised as ripples propagating around the circumference of the droplet.}
	\label{edge mode cartoon}
\end{figure}

In this picture one first realises that the Laughlin state in Eq.~\ref{Laughlin state} describes a circular droplet of fluid of radius $R=\ell_B\sqrt{2\beta N}$ which is both uniformly dense and incompressible.
As such, the only low-energy excitations we can form are area-preserving distortions of the droplet.
These are our edge modes; they are waves which encircle the droplet.
The first few modes are shown in Fig.~\ref{edge mode cartoon}, with the $n^\tr{th}$ edge mode, having a wavelength equal to $\frac{2\pi R}{n}$.
These modes are analogous to the phonons in a lattice in that they are periodic modes of our system which distort the degrees of freedom at each point.
In the same way that a system can then contain multiple phonons with the same wave-number, thus making them bosonic objects, our quantum Hall system is also able to support multiple edge excitations of the same $n$, and so our edge modes are also bosonic.

It is then possible to derive the behaviour of these modes in the presence of some confining potential in a semi-classical manner as in Wen's hydrodynamic formulation\cite{wen1992theory}.
This argument proceeds by considering the classical energy of a charged fluid in the presence of an electric field and then canonically quantises the resulting Hamiltonian.
The result is the chiral linear Luttinger liquid, meaning that these waves propagate around the circumference of the droplet in one direction only.
More explicitly, the statement that the Laughlin edge is a chiral linear Luttinger liquid means that the edge corresponds to a free conformal field theory of bosons.

It should be noted that the cartoon pictures in Fig.~\ref{edge mode cartoon} do not represent the probability density of the edge state wavefunction itself but are instead a representation of the density-density correlation function around the droplet's circumference.
Furthermore, the picture is more complicated in the Moore-Read case, where there also exists a fermionic branch of edge excitations which cannot be visualised in this way.
Nevertheless, the low-energy dynamics of these modes corresponds, as in the Laughlin case, to a free conformal field theory, albeit a product of bosons \textit{and} fermions in this case.

\subsubsection{Parent Hamiltonians}

Given a trial state, it is possible in many cases to work backwards and explicitly construct a parent Hamiltonian, $\ham_\tr{Parent}$, for which the trial state is the exact ground state and the edge states low-energy eigenstates.
In the Laughlin case this ideal Hamiltonian comprises ultra short-range interactions consisting of only a finite number of non-zero \textit{Haldane pseudopotentials}\cite{haldane1985finite}, which are all chosen to be positive (corresponding to repulsive interactions).
A rapid review of these special interactions is given in Appendix \ref{Haldane pseudopotentials}.
For more exotic states the interactions are more complicated; the $\mathbbm{Z}_k$ Read-Rezayi series (where $k=2$ corresponds to Moore-Read) are produced by a set of $(k+1)$-body interactions\cite{read1999beyond}.
Within this parent Hamiltonian we also add weak quadratic confinement in the radial direction, i.e, $U(|\bm{r}|) = \frac{1}{2}U_0\bm{r}^2$, where $U$ must be smaller than $V$ to ensure that the system remains gapped.
This specific choice to make $U$ quadratic is made to ensure that the edge spectrum is exactly linear, matching Wen's Luttinger liquid theory from the outset.

We may now start to think more clearly about the original Hamiltonian involving generic interactions and confinement.
To do so, we split it up into
	\eq{\ham = \ham_\tr{Parent} + \delta\ham}
where $\delta\ham$ is defined by this equation; it is difference between the true interactions and the idealised interactions or the deviation of the true, anharmonic confinement of our the system from the quadratic confinement we have imposed.
Recalling then that these trial states are of interest because they approximate true systems quite accurately, we may take $\delta\ham$ to be small and consider it as a perturbation to $\ham_\tr{Parent}$.
The subsequent diagonalisation of $\delta\ham$ is the subject of this paper.

\subsection{Effective Descriptions}
\label{effective descriptions}

\subsubsection{General Construction}

The trial states we will work with can be written as correlation functions of operators from a chiral conformal field theory (CFT)\cite{francesco2012conformal, ginsparg1988applied, read1999beyond, moore1991nonabelions, dubail2012edge}.
These CFTs are made up of two \textit{sectors}, $\tr{CFT}_\tr{U(1)}\otimes\tr{CFT}_\chi$, denoted the \textit{charge sector} and the \textit{statistics sector} respectively.
In this way an individual particle at position $z$ is represented by an operator, $\ver(z)$, which can be decomposed into two parts,
	\eq{\ver(z) = :e^{i\sqrt\beta\varphi(z)}:\chi(z)}
where this first term is the vertex operator of a free massless Bose field, $\varphi(z)$, from the U(1) charge sector and the second term, $\chi(z)$, is from the statistics sector, $\tr{CFT}_\chi$.
Note that the notation $:X:$ refers to the normal ordering of the operator $X$.
Furthermore, the presence of the vertex operator gives our particle a U(1) charge of $\sqrt\beta$.

Using these operators one can construct a quantum Hall wavefunction at filling fraction $\nu=1/\beta$ as the correlation function
	\eq{\Psi_{\bra{v}}(\bm{z}) = \bra{v}c_\beta^{N\vd}\ver^\vd(z_1)\cdots\ver^\vd(z_N)\ket{0} \label{statedef}}
where $\bm{z}=\{z_1,\hdots,z_N\}$ are the positions of the particles and $\ket{0}$ is the vacuum of the full CFT with zero charge.
The operator $c_\beta^N$ is the \textit{background charge} and is an operator with a U(1) charge of $-N\sqrt\beta$.
Its presence gives the correlator a net U(1) charge of zero, without which the correlator must vanish.

Finally, the out-state, $\bra{v}$, is a state in the full CFT which defines an individual edge excitation with the vacuum, $\bra{0}$, corresponding to the ground state.
Effectively, $\Psi$ can be considered as some linear mapping from a CFT state to a physical wavefunction,
	\eq{\Psi : \ket{v} \mapsto \Psi_{\bra{v}}(\bm{z}).}
The CFT states have a well-defined quantum number, $\Delta L$, which, in the CFT language, is called the \textit{conformal dimension}.
The eigenoperator associated with $\Delta L$ is $L_0$, the $0^\tr{th}$ mode of the \textit{Virasoro algebra}, which generates dilations and is proportional to the chiral CFT's Hamiltonian (as another example, the $-1^\tr{st}$ mode of the Virasoro algebra, $L_{-1}$, generates translations).
The vacuum, $\ket{0}$, has a conformal dimension of $\Delta L=0$.
This quantum number also has an interpretation in the quantum Hall language as the angular momentum of a particular state relative to the vacuum state, $\Psi_{\bra{0}}$.

\subsubsection{Laughlin State}

The Laughlin state has a trivial statistics sector, i.e, $\chi=\mathbbm{1}$, making its CFT solely that of the free boson, $\tr{CFT}_{\tr{U(1)}}$.
This CFT contains only the field
	\eq{\varphi(z) = \varphi_0 - ia_0\ln(z) + i\sum_{n\neq0}\frac{a_n}{n}z^{-n}
			\label{bosonic mode expansion}}
where the modes of the field satisfy the commutation relations
	\eq{[a_n,a_{-m}] = n\delta_{n,m}, \qquad [\varphi_0,a_0] = i.}
All other commutators are trivial.
We then define the vacuum of the theory, $\ket{0}$, as the state which is annihilated by all the positive modes,
	\eq{a_n\ket{0} = 0 \qquad \forall\;n>0.}
We define the background charge in terms of these modes as
	\eq{c_\beta^N = e^{-iN\sqrt\beta\varphi_0}.}

The particle operator for this case is then simply the vertex operator.
Given that the operator product expansion (OPE) of two bosonic field has the form $\varphi(z)\varphi(w)\sim -\ln(z-w)$ we find, using the Baker-Campbell-Hausdorff formula, that
	\eq{\ver(z)\ver(w) = (z-w)^\beta:e^{i\sqrt\beta\lr{\varphi(z)+\varphi(w)}}:.}
From this it is relatively straightforward to see that the ground state, i.e, the state where $\bra{v}=\bra{0}$, has the form
	\eq{\Psi_{\bra{0}}(\bm{z}) = \prod_{i<j}(z_i-z_j)^\beta.}
This is almost as expected for the Laughlin state but with the omission of the Gaussian factor we see in Eq.~\ref{Laughlin state}.
This is due to our choice of background charge, which is equivalent to placing a particle with U(1) charge $-N\sqrt\beta$ at infinity.
It is instead possible to spread this charge over the droplet, at which point the Gaussian factors are recovered, but this is slightly more complicated\cite{moore1991nonabelions, hansson2007composite}.
Therefore, we instead use this simpler version and include the Gaussian factors as part of the integration measure for the problem, which we define as
	\eq{\bbrakket{\Psi_{\bra{v}}}{\Psi_{\bra{w}}} = \int\De\bm{z}\bar\Psi_{\bra{v}}\Psi_{\bra{w}} \label{integration measure}}
where
	\eq{\De\bm{z} = \prod_i\lrs{\de^2z_i\exp\lr{-\frac{|z_i|^2}{2\ell_B^2}}}.}

Excited states, $\bra{v}$, are generated by applying the positive modes $a_n$ to the out-state vacuum, where each $a_n$ raises the conformal dimension of the state by $n$.
Specifically we can define the states
	\eq{\bra{\blam} = \bra{0}\prod_{n\in\blam}a_n,\label{Laughlin edge states}}
where $\blam=\{\lambda_1,\lambda_2,\hdots\}$ is a semi-ordered ($\lambda_1\ge\lambda_2\ge\hdots$) set of positive integers.
The subsequent wavefunction, $\Psi_{\bra{\blam}}$, is the Laughlin state multiplied by a symmetric polynomial,
	\eq{\Psi_{\bra{\blam}}(\bm{z}) = P_\blam\prod_{i<j}(z_i-z_j)^\beta}
where $P_\blam$ is a product of \textit{power sums},
	\eq{P_\blam = \prod_{n\in\blam}p_n,\qquad\qquad p_n=\sqrt\beta\sum_iz_i^n}
(note that we do not normalise the $z_i$ by factors of the radius, $R$, as is sometimes the convention).
As such, we can now see how the conformal dimension of the state corresponds to the added angular momentum.
If we recall that the power on any $z_i$ is the angular momentum of that particle, we see that this polynomial adds $\sum_i\lambda_i$ units of angular momentum to the ground state, and this is exactly the conformal dimension of the state $\ket{\blam}$.

\subsubsection{Moore-Read State}

The statistics sector for the Moore-Read wavefunction is that of a free Majorana fermion whose field, $\chi(z)=\psi(z)$, has an OPE of the form
	\eq{\psi(z)\psi(w) \sim \frac{1}{z-w}}
and admits a mode expansion of the form
	\eq{\psi(z) = \sum_{n\in\mathbbm{Z}+\frac{1}{2}}\psi_nz^{-n-1/2}.}
These modes satisfy the anti-commutation relation
	\eq{\{\psi_n,\psi_{-m}\} = \delta_{n,m}}
and yield a vacuum, $\ket{0}$, which is annihilated by the positive fermionic modes, $\psi_n$ for $n>0$ (in addition to $\ket{0}$ being annihilated by the positive modes of the bosonic field; $\ket{0} = \ket{0}_{\tr{U}(1)}\otimes\ket{0}_\psi$).
The background charge is unchanged from the Laughlin case.

Before we construct the trial wavefunctions which follow from this CFT it is worth noting that the fermionic CFT contains a parity symmetry.
This symmetry forces the correlation function of an odd number of fermionic fields to vanish, and so the construction differs slightly between odd and even particle numbers.
We shall focus initially on the even case.
In this case the ground state is of the form
	\eq{\Psi_{\bra{0}}(\bm{z}) = \tr{Pf}\lr{\frac{1}{z_i-z_j}}\prod_{i<j}(z_i-z_j)^\beta.
		\label{MR definition}}
where this extra term, $\tr{Pf}(\hdots)$, arising from the contraction of the fermionic fields is called the \textit{Pfaffian}, and is an antisymmetrised sum over all products of the fractions $\frac{1}{z_i-z_j}$,
	\eq{\tr{Pf}\lr{\frac{1}{z_i-z_j}} = \mathbbm{A}\lr{\frac{1}{z_1-z_2}
				\frac{1}{z_3-z_4}\cdots\frac{1}{z_{N-1}-z_N}}}
where $\mathbbm{A}$ refers to the antisymmetrisation over all the indices $1,\hdots,N$.
Once again, the form of the wavefunction Eq.~\ref{MR definition} omits the necessity Gaussian factors which we once again place within the integration measure, whose form is exactly equivalent to Eq.~\ref{integration measure}.

Consistent with the Laughlin state, we excite edge modes by applying the positive modes of the fields in our CFT on the vacuum.
In this case, we have two branches of excitations, one of which is an exact replication of the bosonic excitations seen for the Laughlin and another from the fermionic field.
This second branch is more restricted than for the bosons given that the states must obey parity symmetry and possess a fermionic exclusion principle.
As such, general states have the form
	\eq{\bra{\blam\;;\;\bmu} = \bra{0}\prod_{n\in\blam}a_n\prod_{l\in\bmu}\psi_l
			\label{Moore-Read edge states}}
where $\blam$ is once again a semi-ordered set of positive integers whilst $\bmu$ is an ordered ($\mu_1>\mu_2>\hdots$) set of positive half-integers ($\mu_i\in\left\{\frac{1}{2},\frac{3}{2},\frac{5}{2},\hdots\right\}$).
Parity symmetry then forces the number of elements within the set to be even.
As the $\psi_l$ anti-commute we must also enforce an ordering on this product and we choose to order the $\psi_l$ with the smallest $l$ at the left-most position, i.e,
	\eq{\bra{\emptyset\;;\;\frac{3}{2},\frac{1}{2}} = \bra{0}\psi_{\frac{1}{2}}\psi_{\frac{3}{2}}.}
Despite the extra complexity of these states, they once again raise the angular momentum by an amount equal to the conformal dimension of the state, which in this case is
	\eq{\Delta L = \sum_i\lambda_i + \sum_j\mu_j.}

Finally, we consider the case when $N$ is odd.
The picture is almost identical but, due to parity symmetry, we must be careful to ensure that the total number of fermionic operators within any given correlator is even.
As such, the edge states from Eq.~\ref{Moore-Read edge states} are identical except for the condition that $\bmu$ must be a set containing an \textit{odd} number of elements.
This even holds for the ground state, which is instead defined by the CFT state $\bra{0}\psi_{\frac{1}{2}}$.
Therefore, note also that the angular momentum added by state $\bra{v}$ is now the conformal dimension of $\bra{v}$ less the conformal dimension of this ground state (whose conformal dimension is $\frac{1}{2}$).

\section{Effective Hamiltonians}
\label{hamiltonians}

\subsection{The Effective Hamiltonian}

We are now in a position to use the powerful language of CFT to generate effective low-energy theories for quantum Hall edges.
As we have already discussed, we do so by considering the problem as one of degenerate perturbation theory.
In the construction we have just introduced our physical states are labelled by auxiliary states in the CFT $\Psi_{\bra{v}}$.
The state $\bra{v}$ is such that it describes a state with $\Delta L$ units of angular momentum with respect to the ground state.
The parent Hamiltonian, with its parabolic confinement, is then such that the energies of these states are linear in this added angular momentum, $E_v\propto\Delta L$.

However, these subspaces at a given $\Delta L$ will be degenerate.
For example, in the Laughlin case at $\Delta L=2$ there are two states, $\Psi_{\bra{2}}$ and $\Psi_{\bra{1,1}}$.
Once we impose our perturbation, $\delta\ham$ on the system, this degeneracy will in general break, mixing the two states,
	\eq{\delta\ham\Psi_{\bra{v}} = \sum_wH_{v,w}\Psi_{\bra{w}}. \label{perturbation mix}}
However, given the linearity of the description of these wavefunctions in terms of CFT (i.e, $\alpha\Psi_{\bra{v}} + \beta\Psi_{\bra{w}} = \Psi_{\bra{v}\alpha+\bra{w}\beta}$), we in fact have that
	\eq{\delta\ham\Psi_{\bra{v}} = \Psi_{\bra{v'}}\quad\tr{where}\quad
				\bra{v'} = \sum_w\bra{w}H_{v,w}.}
As such, there is an operator $H$ which is the image of $\delta\ham$ under a linear mapping from the physical space of states to the CFT which reproduces the mixing of the real states in the CFT language, i,e,
	\eq{\delta\ham\Psi_{\bra{v}} = \Psi_{\bra{v}H}.}
This equation defines $H$.

In what follows we will consider the constraints imposed upon $H$ by the symmetries of $\delta\ham$.
As usual in quantum mechanics, the symmetries of the Hamiltonian will be expressed as a vanishing commutation relation with some operator, $\mathcal{B}$, which encodes the particular symmetry.
Consider then that this operator also has a mapping to the CFT,
	\eq{\mathcal{B}\Psi_{\bra{v}} = \Psi_{\bra{v}B}.}
In this way, the symmetry of $\delta\ham$ can be simply mapped to a symmetry of $H$,
	\eq{[\mathcal{B},\delta\ham]=0 \quad\mapsto\quad [H,B]=0.}
This procedure allows us to impose strong constraints on the form of $H$.

It is worth noting a key consequence of the fact that the mapping is linear.
Consider, for example that we perturb the trial wavefunction by two perturbations, $\delta\ham = \delta\ham_1 + \delta\ham_2$.
Given that our mapping to the CFT language is linear, each of these perturbations admits its own effective description and the two simply add,
	\eq{\delta\ham = \delta\ham_1 + \delta\ham_2 \quad\mapsto\quad H = H_1 + H_2.}
Now consider that $\delta\ham$ commutes with a set of operators $\{\mathcal{B}_1,\hdots,\mathcal{B}_n\}$ but that $\delta\ham_2$ also has one extra symmetry, $\mathcal{B}_{n+1}$.
The former statement implies that $H$ commutes with each of $B_1$ to $B_n$.
However, $\delta\ham_2$ also commutes with $\mathcal{B}_{n+1}$ and, given that the mapping to the CFT language is linear, it must also be the case that $[H_2,B_{n+1}]=0$.
Therefore, the individual effective Hamiltonian satisfies symmetries that the whole might not.

This may seem like an obvious point but it is of crucial importance.
Consider for example that $\delta\ham_1$ is some confinement imposed on the system and $\delta\ham_2$ corresponds to an interaction.
In this case, $\delta\ham_2$ possesses an extra symmetry to $\delta\ham_1$, that of translational invariance, and this imposes extra constraints on the form of $H_2$.
However, because the mapping $\delta\ham\to H$ is linear, the only part of the effective Hamiltonian that knows anything about the form of the interactions is $H_2$.
Therefore, even when the generic perturbation to the system as a whole, $\delta\ham$, does not possess translational symmetry, the fact that our mapping is linear means that we are still able to make strong statements about all the contributions to $H$ arising from interactions.

\subsection{Preliminary Example - The Harmonic Case}

We begin with a simple example where the mapping $\delta\ham\mapsto H$ can be performed exactly.
In general, this will not be possible but this example provides a taste of the machinery involved in the subsequent calculations.

The perturbation we will consider is a harmonic confinement imposed upon the droplet.
In the first quantised language this perturbation has the form
	\eq{\delta\ham = U_0\sum_{i=1}^N\left|\frac{z_i}{R}\right|^2.}
We now wish to consider the effect of $\delta\ham$ acting upon our wavefunction $\Psi_{\bra{v}}$.
At first glance this does not appear possible as the wavefunction $\Psi_{\bra{v}}$ can only be holomorphic but the interaction contains $\bar z_i$ terms.
We are saved in this instance by a procedure known as \textit{projection to the lowest Landau level}.
Effectively, we need only consider the matrix elements of $\delta\ham$ within our subspace of states given by Eq.~\ref{statedef} and where the integration measure is given by Eq.~\ref{integration measure}.
In this case we can replace any $\bar z_i$ in the integral with the differential operator $-2\ell_B^2\partial_i$ acting on the exponential factors inside the integration measure.
Thus, following integration by parts we find that
	\eq{\bar z_i \mapsto 2\ell_B^2\partial_i. \label{projection to LLL}}

In this way, we can reformulate our perturbation as a differential operator.
Recalling that $R=\ell_B\sqrt{2\beta N}$ we find,
	\eq{\delta\ham = \frac{U_0}{\beta} + \frac{U_0}{\beta N}\sum_{i=1}^Nz_i\partial_i}
(where the constant term arises from the action of $\partial_i$ on $z_i$ itself).
Thus, we find a constant energy shift plus an extra term which we can map into the CFT using a \textit{Ward identity}\cite{francesco2012conformal}.
Using the identity
	\eq{[L_0,\ver(z)] = \lr{z\partial + \frac{\beta}{2}}\ver(z)}
we are able to map $\delta\ham$ into the CFT as
	\eq{\delta\ham\bra{v}c_\beta^N\ver(z_1)\hdots\ket{0} = \bra{v}c_\beta^N\tilde H\ver(z_1)\hdots\ket{0}}
where this operator $\tilde H$ is
	\eq{\tilde H = \frac{U_0}{\beta} + \frac{U_0}{\beta N}\lr{L_0 - \frac{\beta N}{2}}.}
Therefore, to find the Hamiltonian we simply need to commute $\tilde H$ through the background charge, which takes $a_0\to a_0+N\sqrt\beta$ and therefore gives
	\eq{H = \frac{U_0}{\beta N}\lr{L_0 + a_0N\sqrt\beta} + U_0\frac{\beta(N-1)+2}{2\beta}.}

Thus, we have our first effective Hamiltonian.
In the space of neutral edge states (i.e, $a_0\ket{v}=0$) this simply reduces to a linear model,
	\eq{H = \frac{U_0}{\beta N}L_0 + \tr{const}}
where $L_0$ simply counts the conformal dimension, $\Delta L$, of the state $\ket{v}$, and so
	\eq{E = \frac{U_0}{\beta N} \Delta L + \tr{const}.}
This also proves that any perturbing confinement to the quantum Hall system which is quadratic gives only a linear contribution to the spectrum.

Note that a similar mapping is also possible for a quadratic interaction,
	\eq{\delta\ham = V_0\sum_{i\neq j}\left|\frac{z_i-z_j}{2\ell_B}\right|^2.}
We find that
	\eq{H = NV_0\lr{L_0 - a_1a_{-1} - \frac{1}{N\sqrt\beta}a_1L_{-1}} + \tr{const}}
in the sector where the states have zero U(1) charge (i.e, $a_0=0$).
The derivation of this result is presented in Appendix \ref{Harmonic Interactions}.

\subsection{Symmetries of the Effective Hamiltonian}

A priori, we know nothing about the form of $H$ but we shall attempt to constrain it using a simple symmetry analysis.
In what follows we shall use a local field theory description for $H$ and then map the symmetries of the perturbation, $\delta\ham$, into the CFT language in order to find symmetry constraints on the terms in $H$.

\subsubsection{A Local Field Theory}

We begin by noting that our effective Hamiltonian operator, $H$, is a CFT operator which acts only within the space of edge states and so is supported along the edge.
We then conjecture this Hamiltonian should be local, following from the work of Dubail, Read and Rezayi\cite{dubail2012edge} who emphasised the importance of local field theories in the description of quantum Hall states.
To motivate this idea we appeal to the generalised screening hypothesis, which states that correlations within the bulk of quantum Hall droplets decay exponentially\cite{laughlin1983anomalous, girvin1999quantum}.
For the Laughlin state, this can be understood in terms of the plasma analogy, which re-imagines the probability distribution of particles in the ground state as the partition function of a two-dimensional plasma of charge-$\beta$ particles.
Numerical work\cite{caillol1982monte} suggests that this plasma is in a screening phase when $\beta<\beta_c$ for a critical value of $\beta_c\simeq 65$.
Hence, the interactions between particles in this plasma decay exponentially and one expects the relevant physics to be local.
A generalised screening argument also exists for the Moore-Read state which, whilst much more in-depth, also concludes that bulk correlations decay exponentially\cite{bonderson2011plasma}.

Thus, we are conjecturing that our effective Hamiltonian, which by its definition is some operator supported along the edge of our droplet, should be local.
This means that we take $H$ to be of the form
	\eq{H = \int\de x\sum_ah_a(x, \delta\ham)\Phi_a(x) \label{local hamiltonian}}
where $x$ is the one-dimensional coordinate encircling the surface of the droplet and $\Phi_a(x)$ are local operators made up of the fields $\varphi(x)$ and $\chi(x)$ and which have associated coupling constants $h_a$ that depend a priori on the position, $x$, and the perturbation, $\delta\ham$.
A non-local form would be an integral over multiple edge coordinates with local Hamiltonian densities depending in some non-trivial manner on each coordinate, thus coupling well-separated regions of space along our edge.

The mapping between this edge coordinate $x$ and a complex planar coordinate $z$ is given by $z = re^{ix/R}$ for any $r$ (but physically we consider $r\simeq R$) and where $R$ is the droplet's radius.
Working with this planar coordinate proves to be a large simplification.
If we make this change of variables we find that
	\eq{H = \sum_a\oint\frac{\de z}{2\pi i}h_a(z, \delta\ham)\lr{\frac{z}{R}}^{d_a-1}\Phi_a(z) \label{local hamiltonian fields}}
where $d_a$ is the scaling dimension of the field $\Phi_a$ (and we have also absorbed some constant factors into $h_a$).
Given that we are interested in cases where the number of particles is finite but still large, we see that the effective Hamiltonian is an expansion in $\frac{1}{R}$ where $R$ is large.
As such, we may restrict ourselves to considering only contributions with a small scaling dimension, $d_a$, and still hope to gain an accurate picture for relatively large system sizes.
Furthermore, in the scaling limit, for which $R\to\infty$, the behaviour is given simply by the term with the lowest scaling dimension, $d_a$.

It should be stressed that this locality conjecture is not a rigorous constraint on $H$.
We have motivated it here on the idea that the bulk physics is local though exactly how this should transfer to the form of $H$ is not fully understood (though we provide some further evidence in the integer quantum Hall effect in a future publication\cite{bondesanFUTUREinteger}).
Therefore, we provide supporting numerical evidence in section \ref{numerics} which further substantiates that this local description is very accurate for at least short-range interactions.
Nevertheless, the locality conjecture may incur some loss of generality when the perturbation we add to the Hamiltonian describes some long-range interactions, in which case one might no longer expect a local field theory to provide an ample description of the dynamics.
However, without the powerful simplification that this conjecture imposes on the theory, it would be extremely difficult to make significant progress.

\subsubsection{Number Conservation}

We now move onto rigorous constraints on our Hamiltonian, of which we shall consider three.
The first is the conservation of total number of particles in the system.
This is conservation of U(1) charge, as each particle possesses a charge of $\sqrt\beta$.
The operator which counts this charge is $a_0$ and therefore, the particle number in the CFT language is $\hat N = a_0/\sqrt\beta$.
As such, the Hamiltonian must commute with $a_0$,
	\eq{[H,a_0] = 0.}

The consequence of this is relatively simple and means that our Hamiltonian must obey the same underlying U(1) symmetry of the free boson CFT.
For the field this symmetry manifests itself as a shift to the field under the action of the U(1) generator
	\eq{\varphi(z) \to \varphi(z) + \delta\varphi_0.}
As such, the individual operators $\Phi_a(z)$ can only involve $\varphi(z)$ as a derivative, i.e, $\partial^n\varphi(z)$ for $n>0$.
Note that there is no constraint on the statistics sector due to this conservation law.

\subsubsection{Rotational Invariance}

We will also consider perturbations $\delta\ham$ which are rotationally invariant.
Once again, this is a very reasonable constraint for interactions, though perhaps more restrictive for confining potentials.
On a mathematical level, it is relatively simple to derive the concomitant commutation relation for this symmetry by noting that the Hamiltonian should leave the total amount of angular momentum in the system invariant.
Recalling that the angular momentum relative to the ground state is equal to the conformal dimension of the state $\bra{v}$, and this is measured by the operator $L_0$, we therefore have that
	\eq{[H,L_0] = 0.}

However, once again, it is perhaps simpler to consider this constraint from a more physical perspective.
Our droplet is a disc, and this system is invariant under rotations.
Our edge is the circle at the edge of this disc, and the rotation of the bulk coordinate corresponds to a translation of the edge coordinate.
Therefore, our edge Hamiltonian must be translationally invariant.
This has a simple consequence that the coupling coefficients $h_a$ which we introduced in Eq.~\ref{local hamiltonian} must be independent of $x$.

\subsubsection{Translational Invariance}

The final bulk symmetry we may consider is two-dimensional translational invariance.
Of course, this symmetry is not applicable to confining potentials but it does provide very strong constraints on the forms of field theories which describe interaction perturbations.
Such perturbations will be translationally invariant within the bulk, meaning that the perturbation commutes with the generator of translations,
	\eq{\left[\delta\ham\;,\; \sum_i\partial_i\right] = 0. \label{translational invariance}}
where $\partial_i=\frac{\partial}{\partial z_i}$.
Thus, if we can map this generator into our conformal field theory, i.e,
	\eq{\sum_i\partial_i\Psi_{\bra{v}} = \Psi_{\bra{v}\mathcal{D}}, \label{D definition}}
then we know that our effective Hamiltonian, $H$, must commute with $\mathcal{D}$.

\begin{figure}[ht!]
	\centering
	\begin{tikzpicture}
		\draw [white] (-1.5,-1.5)--(1.5,1.5);
		\fill [orange,opacity=0.5,domain=0:360,samples=100,variable=\t] plot ({\t}:{1+0.1*cos(9*\t)});
		\draw [dashed] (0,0) circle (1);
		\draw [ultra thick, ->] (-90:1) arc (-90:90:1) node [anchor=south west] {$\delta\ham$};
		\fill (0,0) circle (0.05);
	\end{tikzpicture}
	$\qquad$
	\begin{tikzpicture}
		\def\dx{-0.2}
		\def\dy{-0.2}
		\draw [white] (-1.5+\dx,-1.5+\dy)--(1.5+\dx,1.5+\dy);
		\fill [orange,opacity=0.5,domain=0:360,samples=100,variable=\t] plot ({\t}:{1+0.1*cos(9*\t)});
		\draw [dashed] (\dx,\dy) circle (1);
		\draw [ultra thick, ->] (-90:1) arc (-90:90:1) node [anchor=south west] {$\delta\ham$};
		\draw [ultra thick, ->] (\dx,\dy) -- (0,0);
		\fill (\dx,\dy) circle (0.05);
	\end{tikzpicture}
	\caption{On the left we show a droplet with some edge excitation which is evolved by our translationally invariant perturbation, $\delta\ham$.
		Given that $\delta\ham$ is translationally invariant, this evolution will be about the centre of mass, and thus should be the same for the droplet on the right, whose centre of mass is slightly shifted.
		Crucially, $\delta\ham$ cannot move this centre of mass.}
	\label{shifted dynamics}
\end{figure}
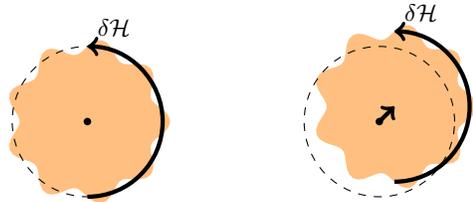

Nevertheless, there is once again a simpler picture to keep in mind.
Consider that our perturbation $\delta\ham$ will induce dynamics on the edge states of our system.
This is shown by another cartoon picture in Fig.~\ref{shifted dynamics}.
On the left of this figure the droplet is centred and the arrow indicates the dynamics induced by $\delta\ham$ (there may be additional dynamics due to the parabolic confinement in $\ham_\tr{parent}$ which we ignore here).
Now consider the right picture, where we shift the position of our droplet.
The dynamics induced by $\delta\ham$, assuming that it is a translationally invariant perturbation, should be identical, but now about a new origin.
Therefore, there must be some decoupling of the centre-of-mass mode from the underlying field theory.
If we reconsider our cartoon pictures for the edge modes in Fig.~\ref{edge mode cartoon} then we see that the $a_n$ edge mode creates $n$ equally sized lobes of charge around the surface of the droplet.
For $n\ge2$ these cancel out, leading to no net shift of the centre-of-mass.
The $a_1$ mode however, is equivalent to a shift of the droplet.
As such, the field theory, $H$, cannot induce any dynamics on this edge mode.
Therefore, a state $\bra{v}$ will have the same energy under $\delta\ham$ as a shifted state $\bra{v}a_1$ and so
	\eq{[H,a_1] = 0. \label{translation commutator 1}}

However, this argument is far from rigorous and is too much of a simplistic treatment which does not capture the full complexity of the situation.
Therefore, we perform the explicit mapping of the symmetry in Eq.~\ref{translational invariance} directly into the CFT language in two different ways, producing two (closely related) consequences for our effective Hamiltonian.

\textbf{Constraint 1} - For the first consequence, we simply map the generator of translations into the edge mode language.
This can be done quite simply given the $L_{-1}$ Virasoro mode, which is the generator of translations in the CFT language.
This operator has the properties that
	\eq{[L_{-1},\ver(z)]=\partial\ver(z), \qquad L_{-1}\ket{0}=0. \label{L-1 action}}
As such, it is simple to reproduce the action of our translation operator on the correlation function via
	\eq{\sum_i\partial_i\Psi_{\bra{v}} = \bra{v}c_\beta^NL_{-1}\ver(z_1)\hdots\ver(z_N)\ket{0}.}
This is one step away from finding the CFT operator $\mathcal{D}$ as defined in Eq.~\ref{D definition}.
We now simply need to move $L_{-1}$ through the background charge.

To do so, we need to better understand the exact form of $L_{-1}$.
This operator is a sum of the Virasoro modes from each CFT,
	\eq{L_{-1} = L_{-1}^{\tr{U(1)}} + L_{-1}^{\chi}}
where
	\eq{L_{-1}^{\tr{U(1)}} = \frac{1}{2}\oint\cint{z}:\lr{i\partial\varphi(z)}^2:}
and $L_{-1}^\chi$ is the $-1$ mode of the stress energy tensor, $T(z)$, of CFT$_\chi$.
Therefore, $L_{-1}^\chi$ moves through the background charge (which depends only on $\varphi_0$) without consequence whereas $L_{-1}^{\tr{U(1)}}$ acquires an extra term.
Specifically, when the field $\varphi(z)$ is conjugated by the background charge it becomes
	\eq{c_\beta^N\varphi(z)c_\beta^{-N} = \varphi(z) - iN\sqrt\beta\ln(z).}
As such, one can conjugate the full $L_{-1}$ operator by the background charge to find that
	\eq{\mathcal{D} = L_{-1} + N\sqrt\beta a_{-1},}
which implies that our effective Hamiltonian has the symmetry
	\eq{\left[H, a_{-1} + \frac{1}{N\sqrt\beta}L_{-1}\right] = 0. \label{translation commutator 2}}

\textbf{Constraint 2} - The second consequence of translational invariance is exactly that in Eq.~\ref{translation commutator 1}, that $[H,a_1]=0$.
To prove this we will consider the hermitian conjugate of the derivation which produced the result in Eq.~\ref{translation commutator 2}.
In that derivation we began with the condition $[\delta\ham, \sum_i\partial_i]=0$.
When considering the matrix elements of this condition,
	\eq{\bbra{\Psi_{\bra{v}}}\left[\delta\ham,\sum_i\partial_i\right]\kket{\Psi_{\bra{w}}} = 0, \label{symmetry matrix def}}
we mapped both $\sum_i\partial_i$ and $\delta\ham$ into the CFT to find that
	\eq{\bbra{\Psi_{\bra{v}}}\left[\delta\ham,\sum_i\partial_i\right]\kket{\Psi_{\bra{w}}} = \bbrakket{\Psi_{\bra{v}}}{\Psi_{\bra{w}[\mathcal{D},H]}} = 0. \label{symmetry matrix}}
Therefore, we claimed that, for this to hold for any states $\bra{v}$ and $\bra{w}$, it must be that $\Psi_{\bra{w}[\mathcal{D},H]}=0$ which implies that $[\mathcal{D},H]=0$.

We now run through the hermitian conjugate of this argument.
Firstly, the perturbation, $\delta\ham$, is hermitian and therefore can act backwards on the bra-state,
	\eq{\bbra{\Psi_{\bra{v}}}\delta\ham\kket{\Psi_{\bra{w}}} = \bbrakket{\Psi_{\bra{v}H}}{\Psi_{\bra{w}}}.}
Therefore, if we can also find a mapping of $\sum_i\partial_i$ on the bra-state, i.e, some $\mathcal{K}$ such that
	\eq{\bbra{\Psi_{\bra{v}}}\sum_i\partial_i\kket{\Psi_{\bra{w}}} = \bbrakket{\Psi_{\bra{v}\mathcal{K}}}{\Psi_{\bra{w}}}.}
then we have another condition on $H$, namely $[H,\mathcal{K}] = 0$.

To find $\mathcal{K}$ we need to find a way to convert the translation operator, $\sum_i\partial_i$, which acts on holomorphic wavefunctions, into some operator which acts only on anti-holomorphic wavefunctions, i.e, the bra-state $\bar\Psi_{\bra{v}}$, which is a function of the $\bar z_i$ instead of the $z_i$.
Recall then the procedure of projection to the lowest Landau level, Eq.~\ref{projection to LLL}, the result of which is that $\bar z_i \equiv 2\ell_B^2\partial_i$ within the lowest Landau level (that $\partial_i\equiv \bar z_i/2\ell_B^2$ can be seen by considering the matrix element of $\partial_i$ in integral form and integrating by parts, taking care to remember the exponential factors in the integration measure, Eq.~\ref{integration measure}).
Therefore, we find that
	\eq{\bbra{\Psi_{\bra{v}}}\sum_i\partial_i\kket{\Psi_{\bra{w}}} = \bbra{\Psi_{\bra{v}}}\sum_i\frac{\bar z_i}{2\ell_B^2}\kket{\Psi_{\bra{w}}}.}
This can be readily mapped to an operator acting on the bra-state's auxiliary state when we remember that the edge mode $a_1$ is simply the polynomial $\sqrt\beta\sum_i\bar z_i$.
Therefore, $\mathcal{K} = a_1/2\ell_B^2\sqrt\beta$, thus confirming that $[H,a_1]=0$ as claimed.

One may well worry about the apparent contradiction between the two constraints in Eqs.~\ref{translation commutator 1} and \ref{translation commutator 2}.
If one takes a Hermitian conjugate of the former, using the fact that $(a_1)^\dagger=a_{-1}$, one does not reproduce the latter if one assumes that the effective Hamiltonian is hermitian.
This is because the effective Hamiltonian is not necessarily hermitian.
As we have defined it, it is simply an operator which reproduces the action of a hermitian perturbation on some auxiliary space.
Therefore, it must have only real eigenvalues.
In fact, in the case of quadratically increasing repulsive interactions which we are able to map exactly into the CFT (see Appendix \ref{Harmonic Interactions}) the resulting Hamiltonian was indeed non-Hermitian.
The cause of this non-hermiticity is that the inner product of two quantum Hall states labelled by different $\bra{v}$ are not necessarily orthogonal.
In fact,
	\eq{\bbrakket{\Psi_{\bra{v}}}{\Psi_{\bra{w}}} = \bra{w}G_N\ket{v}}
where $G_N$ is some operator close to, but not equal to, the identity\cite{dubail2012edge, fern2018structure}.
Therefore, whilst $H$ need not be hermitian by itself, one does require that $HG_N$ be hermitian.

\subsubsection{Confinement vs Interactions}

To summarise the results of the preceding section, we have found the mapping of various symmetries of a general perturbation $\delta\ham$ into our CFT language.
They are as follows:
\begin{enumerate}
	\item Number conservation implies that
		\eq{[H,a_0] = 0,}
	\item Rotational invariance implies that
		\eq{[H,L_0] = 0,}
	\item Translational invariance of $\delta\ham$ implies that
		\alg{[H, a_1] & = 0 \\
				\left[H, a_{-1} + \frac{1}{N\sqrt\beta}L_{-1}\right] & = 0.}
\end{enumerate}
Note that these are general symmetries which apply to $H$ \textit{regardless of our conjecture of locality}.

Now, consider some generic perturbation $\delta\ham$, which we split up into a confining part $\delta\ham_U$, and an interaction term, $\delta\ham_V$.
As discussed previously, these individual perturbations will have mappings into the CFT of the form $H_U$ and $H_V$ which simply add.
In the rotationally symmetric cases we will consider, $H_U$ will then satisfy the first two symmetries, of number conservation and rotational invariance.
However, we will find that we can constrain the form of $H_V$ significantly more given that it must also satisfy the third symmetry, corresponding to bulk translational invariance.

\section{Results}
\label{results}

\subsection{Confinement}

We begin by considering the effect of number conservation and rotational symmetry on the form of effective Hamiltonians.
This is the situation which corresponds to the part of our field theory which describes the effects of confinement on the droplet.
We recall that number conservation forces $H$ to commute with $a_0$, thus forcing $\varphi(z)$ to appear in $H$ only as a derivative and that rotational invariance equates to one-dimensional translations along the edge, removing the possibility for our coupling coefficients, $h_a$, to depend on position, $z$.

Let us consider the effect of these symmetries for the Laughlin and Moore-Read wavefunctions.
In the Laughlin case the statistics sector is trivial ($\chi=\mathbbm{1}$) leaving only a theory made from the bosonic field.
In general the fields $\Phi_a(z)$ have the form
	\eq{\Phi_a(z) = \lr{i\partial\varphi(z)}^{m_1}\lr{i\partial^2\varphi(z)}^{m_2}\cdots}
for non-negative integers $m_1,m_2,\hdots$.
Each term has a scaling dimension $d_a=m_1 + 2m_2 + \hdots$ and it is always assumed that they are normal ordered.
Therefore, we can consider the first few most relevant terms (up to total derivatives) to be
	\alg{H = \oint\cint{z}&\lr{\frac{v}{2}z\lr{i\partial\varphi(z)}^2 + 
				g z^2\lr{i\partial\varphi(z)}^3} \nonumber\\&
		+ \order{R^{-3}}.}
The first of these terms is the usual chiral linear Luttinger liquid term, and by itself would lead to a dispersion $E = v\Delta L$.
The second term is then a scattering term which, for example, might take the $n=2$ mode (recall Fig.~\ref{edge mode cartoon}) and scatter this into two $n=1$ modes.

We may also consider the consequences of these simple symmetries on the generic effective Hamiltonian describing a Moore-Read edge.
In this case the statistics sector is a free fermion CFT with $\chi(z)=\psi(z)$.
Therefore, our general fields have the form
	\eq{\Phi_a(z) = \lr{i\partial\varphi(z)}^{m_1}\cdots\times\lr{\psi(z)}^{k_0}\lr{\partial\psi(z)}^{k_1}\cdots}
where the $m_i$ are once again any non-negative integer but $k_i\in\{0,1\}$ due to fermionic exclusion.
In this case the scaling dimension of a given term is $d_a = \sum_nnm_n + \sum_l\lr{l+\frac{1}{2}}k_l$.
As such, the first few most relevant terms will be
	\alg{H = \oint\cint{z}&\left(-\frac{v_1}{2}z\lr{\psi\partial\psi(z)} + \frac{v_2}{2}z\lr{i\partial\varphi(z)}^2\right.\nonumber\\&
			\left. + g_1 z^2\lr{i\partial\varphi(z)}^3 + g_2 z^2\lr{i\partial\varphi\psi\partial\psi(z)}\right) \nonumber\\&
		+ \order{R^{-3}}.}
In this Hamiltonian the first two terms are once again linear edge velocities which by themselves would simply give us the spectrum $E = v_1\Delta L_\psi + v_2\Delta L_\varphi$, giving the fermionic modes a velocity $v_1$ and the bosonic modes a velocity $v_2$.
We then have two scattering terms, one exactly equivalent to the Laughlin scattering term, which scatters bosonic modes, and one coupling term which scatters a single bosonic mode into two fermions and vice versa.

\subsection{Interacting Laughlin}

We now consider the addition of two-dimensional translational symmetry to the effective Hamiltonian, which will allow us to describe the effects of interactions on the edge dynamics.
We will find that this symmetry is very restrictive, removing the majority of those terms which were present in those effective Hamiltonians describing confinement.
Given these restrictions, we will need to go to rather high order (large $d_a$) to find the major contributing terms.
In doing so we need to find a good basis of operators with which to work and then use those to construct linearly independent Hamiltonians $H_a$ such that
	\eq{H = \sum_ah_aH_a}
where the $H_a$ are individual blocks which satisfy the translational symmetry commutation relations and $h_a(\delta\ham)$ are coefficients which scale as $h_a\sim R^{1-d_a}$ where $d_a$ is the scaling dimension of the leading term (the term with the lowest scaling dimension) in $H_a$.

\subsubsection{A Basis of Fields}

To consider this problem generally, it is useful to generate a basis of terms which can arise within the field theory.
We label these terms with partitions, $\bgam = \{\gamma_1,\gamma_2,\hdots\}$, of integers, $\gamma_i>0$.
These integers refer to the derivatives we have of the field $\varphi(z)$, as defined by terms
	\eq{T_\bgam = \oint\cint{z}z^{|\bgam|-1}\prod_{\gamma_i\in\bgam}i\partial^{\gamma_i}\varphi(z)}
where the factor of $z^{|\bgam|-1}$ for $|\bgam|=\sum_i\gamma_i$ is required by the rotational invariance condition.
These terms have a scaling dimension $d_\bgam = |\bgam|$.
In this way we may generate any general contribution $H_a$ as some linear combination of these $T_\bgam$ with $d_\bgam = d_a$.

Note that the order of $\gamma_i$ in $\bgam$ is unimportant in the definition of $T_\bgam$ as the fields are bosonic and so we take the ordering $\gamma_1\ge\gamma_2\ge\hdots$.
Note further that the basis is over-complete.
To see this, consider $T_{21}$, which we can integrate by parts to give
	\alg{T_{21} = \oint\cint{z}z^2\frac{1}{2}\partial\lr{\lr{i\partial\varphi(z)}^2} = -T_{11} = -2L_0.\label{T21 is T11}}
Here we used the holomorphicity of the stress-energy tensor at infinity, $T(z)=\order{z^{-4}}$ as $z\to\infty$.
Therefore, one must be careful to use a basis of $T_\bgam$ which includes only `unique' $\bgam$ (i.e, linearly independent $T_\bgam$).
At low orders it is sufficient to find these `unique' $\bgam$ simply by the integration by parts procedure described above, though for higher orders it may be necessary to decompose the subsequent terms into bosonic modes and analyse whether the individual matrices are linearly independent.

It is worth noting an apparent oddity in Eq.~\ref{T21 is T11}.
On the left hand side we have $T_{21}$ with scaling dimension 3 and on the right hand side $T_{11}$, whose scaling dimension is 2.
The source of this contradiction is really just notation.
When we construct the Hamiltonian in Eq.~\ref{local hamiltonian} as $H = \int\de x\sum_a h_a\Phi_a(x)$ we do not expect $\lr{\partial\varphi}^2$ to appear at scaling dimension $d_a=3$.
We only expect $\partial^2\varphi\partial\varphi$ at this order (and also $\lr{\partial\varphi}^3$ and $\partial^3\varphi$).
However, we are searching for a convenient basis to use and so it makes little sense to keep careful track of both $T_{21}$ and $T_{11}$ in the effective Hamiltonian if they are simply the same operator.
Therefore, it is more convenient to simply throw away $T_{21}$ and allow $T_{11}$ to appear at both scaling dimension 2 and 3.
A similar reasoning applies at higher orders and this means that any $T_\bgam$ in the final basis can appear at scaling dimension $d_\bgam$ \textit{and any scaling dimension higher}.
This further implies that the coefficients will vary as $h_a\sim cR^{1-d_a} + c'R^{-d_a}+\hdots$ for some constants $c$, $c'$ and so on.

Once we rid ourselves of these extraneous terms we are left with a linearly independent basis up to scaling dimension 7 (or $R^{-6}$) given in table \ref{terms table}.
Note that we do not consider the scaling dimension $d_\bgam = 1$ as the sole term here is $T_1=a_0$ and we work with states $\ket{\blam}$ (recall the definition in Eq.~\ref{Laughlin edge states}) which have U(1) charge of zero (i.e, $a_0\ket{\blam}=0$ for all $\ket{\blam}$).
Therefore, this term and others like it are trivial.

\begin{table}[ht!]
	\begin{tabular}{| c | l |}
		\hline$d_\bgam$ & Unique terms \\\hline\hline
		2 & $T_{11}$ \\
		3 & $T_{111}$ \\
		4 & $T_{22}, T_{1111}$ \\
		5 & $T_{221}, T_{11111}$ \\
		6 & $T_{33}, T_{222}, T_{2211}, T_{111111}$ \\
		7 & $T_{331}, T_{2221}, T_{22111}, T_{1111111}$ \\
		$\vdots$ & \multicolumn{1}{|c|}{$\vdots$} \\ \hline
	\end{tabular}
	\caption{A table of some of the first few linearly independent terms from which $H$ can be constructed.
		Note that the choices made for $\bgam$ are not unique.
		For example, as we have already seen $T_{11}$ and $T_{21}$ are exactly equivalent up to an overall sign and so to keep notation simpler we replace $T_{21}$ with $T_{11}$, though must then allow part of the coefficient of $T_{11}$ to vary as $R^{1-d_{21}}$.}
	\label{terms table}
\end{table}

\subsubsection{Constraining $H$}

We will now restrict to interactions, considering $\delta\ham$ to be translationally invariant, thus implying that each of our effective Hamiltonians, $H_a$, must satisfy the commutation relations Eqs.~\ref{translation commutator 1} and \ref{translation commutator 2}, which we recall to have the form
	\alg{[H_a, a_1] & = 0, \label{trans1}\\
		\left[H_a, a_{-1} + \frac{1}{N\sqrt\beta}L_{-1}\right] & = 0. \label{trans2}}
By inspecting Eq.~\ref{trans2} we note that we will need to expand $H_a$ in powers of $\frac{1}{N\sqrt\beta}$ of the form
	\eq{H_a = H_a^{(0)} + \frac{H_a^{(1)}}{N\sqrt\beta} + \frac{H_a^{(2)}}{\lr{N\sqrt\beta}^2} + \hdots.}
Note here that the expansion in powers of $1/N$ can be compared to our overall expansion of the effective Hamiltonian in powers of $1/R$.
To relate the two we recall that the radius varies as the square root of the particle number, $R=\ell_B\sqrt{2\beta N}$.
Therefore, if $H_a^{(0)}$ is made up from terms of scaling dimension $d_a$ this tells us that $H_a^{(1)}$ can be made only from terms with a scaling dimension $d_a+2$ or smaller.
In general, $H_a^{(n)}$ is a combination of terms with scaling dimension $d_a+2n$ or smaller.
Furthermore, recalling than $h_a$ scales as $R^{1-d_a}$ and given that $R\sim\sqrt{N}$ we note that the overall coefficient in front of any $H_a^{(n)}$ must vary as $\sqrt N^{1-d_a-2n}$.

Using this expansion, we find from Eqs.~\ref{trans1} and \ref{trans2} that the leading order terms in any $H_a$ are those which satisfy the simple relations
	\eq{[H_a^{(0)},a_1] = [H_a^{(0)},a_{-1}] = 0. \label{leading order}}
For example, of the terms listed in table \ref{terms table} those which satisfy both of these constraints are $T_{22}$ and $T_{33}$.
A summary of these leading terms up to a scaling dimension of 9 is provided in table \ref{leading table}.
The majority of these are simple bilinear terms, which will predominantly modify the dispersion of the edge modes (see below), though we will also find at $d_a=9$ that there is a three-body scattering term.

Nevertheless, the leading terms do not tell the whole story and we must generate the sub-leading terms via Eq.~\ref{trans2}, which necessarily generates interesting nonlinear scattering terms.
This ladder of sub-leading corrections are generated by the constraints
	\alg{[H_a^{(n)},a_1] & = 0 \qquad \forall n, \label{ladder 1}\\
		[H_a^{(n)},a_{-1}] & = [L_{-1}, H_a^{(n-1)}] \label{ladder 2}}
with the former equation from Eq.~\ref{trans1} latter arising from Eq.~\ref{trans2}.

\begin{table}[ht!]
	\begin{tabular}{| c | c |}
		\hline$d_a$ & $H_a^{(0)}$ \\\hline\hline
		4 & $T_{22}$ \\
		6 & $T_{33}$ \\
		8 & $T_{44}$ \\
		9 & $T_{333}+T_{332}-\frac{1}{3}T_{222}$ \\
		$\vdots$ & $\vdots$ \\ \hline
	\end{tabular}
	\caption{The first few leading contributions to the effective Hamiltonian for the Laughlin case which commute with both $a_1$ and $a_{-1}$ and whose effects are suppressed by factors of $R^{1-d_a}$.
		We note that the first three are bilinears in the bosonic modes, which means that their primary effect is on the dispersion of edge modes.
		The final example at $d_a=9$ is a trilinear, which corresponds to some three-body scattering of bosonic modes.}
	\label{leading table}
\end{table}

\subsubsection{The Leading Contributions}

Let us consider this in greater detail for the least irrelevant term with leading contribution $T_{22}$.
To see that this satisfies the leading order requirements for a translationally invariant effective Hamiltonian, Eq.~\ref{leading order}, consider that in terms of bosonic modes this term has the form
	\eq{T_{22} = \oint\cint{z}z^3\lr{i\partial^2\varphi}^2 = -2\sum_{n>0}(n^2-1)a_{-n}a_n.}
Therefore, it is clear that the $n=\pm1$ modes are absent from this term.
We then proceed to apply Eqs.~\ref{ladder 1} and \ref{ladder 2} to produce the sub-leading terms $H_a^{(n)}$.
We first consider Eq.~\ref{ladder 1} in the context of the individual fields,
	\eq{[i\partial^n\varphi(z), a_1] = -\delta_{n,1}.}
Therefore, in order to satisfy this condition, we must have that $n>1$, and so we only consider $T_\bgam$ where $\bgam$ contains integers greater than or equal to 2 such as $T_{222}$, $T_{33}$, $T_{2222}$ and so on.

We then consider Eq.~\ref{ladder 2}, recalling that $L_{-1}$ acts to differentiate the fields of the CFT\cite{francesco2012conformal}, and so this constraint becomes
	\eq{[H_{22}^{(1)},a_{-1}] = \oint\cint{z}z^3\partial\lr{\lr{i\partial^2\varphi}^2}. \label{22 commutation}}
We must now consider which terms might appear in $H_{22}^{(1)}$.
Recall that by scaling arguments the pool of terms which might appear is restricted to those with a scaling dimension $d_{22}+2=6$ or smaller.
This therefore includes $T_{22}, T_{33}$ or $T_{222}$.
A priori, any of these might appear in $H_{22}^{(1)}$ but, given that $T_{22}$ and $T_{33}$ commute with $a_{-1}$ their coefficients are unconstrained by this equation.
Initially, this appears worrying as it suggests that
	\eq{H_{22}^{(1)} = \alpha T_{22} + \beta T_{33} + \gamma T_{222}}
where we are unable to say anything about $\alpha$ and $\beta$.
However, in practice this is not a problem as $T_{22}$ and $T_{33}$ are also permissible $H_a^{(0)}$ terms, exactly because they commute with $a_{-1}$.
Therefore, whatever the values of $\alpha$ and $\beta$, these coefficients can be combined with $h_{22}$ and $h_{33}$, thus yielding no new coefficients.
This argument is indicative of a more general principle.
When considering what terms might appear in any $H_a^{(n)}$ where $n>0$ one should first remove any terms which commute with $a_{-1}$ and therefore produce their own $H_{a'}^{(0)}$ to avoid adding unnecessary extra coefficients.

Therefore, returning to our example, we note that $H_{22}^{(1)}$ should be made up of only $T_{222}$.
To find how much of this term is present, consider once again the commutation of the mode with the field, which in this case is
	\eq{[i\partial^n\varphi(z), a_{-1}] = \partial^{n-1}(z^{-2}).}
Therefore, if we state that $H_{22}^{(1)} = \alpha T_{222}$ then we find by evaluating Eq.~\ref{22 commutation} that
	\eq{\alpha = \frac{1}{2}.}
Thus,
	\eq{H_{22} = T_{22} + \frac{1}{2N\sqrt\beta} T_{222} + \hdots.}
In fact, this procedure can be continued easily to all orders, with the result that
	\eq{\boxed{H_{22} = T_{22} + \sum_{n=1}^\infty\frac{8}{(N\sqrt\beta)^n}\frac{(2n+1)!!}{(2n+4)!!}T_{2^n}} \label{H22}}
where $2^n$ refers to the partition containing $n$ copies of 2.

We can also consider this picture for the $H_a$ whose leading contribution is $T_{33}$.
For this root the first sub-leading correction as calculated using Eq.~\ref{ladder 2} has the form
	\eq{H_{33} = T_{33} + \frac{5}{2N\sqrt\beta}T_{332} - \frac{15}{2N\sqrt\beta}T_{222} + \hdots.}
In general, this contribution is made from the terms $T_{332^{n-2}}$ and $T_{2^n}$ of the form
	\eq{\boxed{H_{33} = T_{33} + \sum_{n=1}^\infty\lr{\frac{\alpha_n}{(N\sqrt\beta)^n}T_{332^n} + \frac{\beta_n}{(N\sqrt\beta)^n}T_{2^{n+2}}},}}
where the coefficients, $\alpha_n$ and $\beta_n$, satisfying the recursive relation
	\alg{\alpha_n & = \frac{2n+3}{2n}\alpha_n, \\
		\beta_n & = \frac{2n+1}{2n+4}\beta_{n-1} - \frac{6(2n+1)}{(n+1)(n+2)}\alpha_n,}
for all $n>0$ and where $(\alpha_0,\beta_0) = (1,0)$.
A similar procedure can be repeated to fix all the $H_a$, leaving only $h_a$ as free parameters which depend on the bulk interactions and cannot be fixed by symmetry only.

\subsection{Interacting Moore-Read}

Effective Hamiltonians for the Moore-Read state follow an extremely similar pattern to that we have seen for the Laughlin state.
Once again, the imposition of bulk translational symmetry on the effective Hamiltonian prompts us to search for individual, independent contributions, $H_a = H_a^{(0)} + \frac{1}{N\sqrt\beta}H_a^{(1)}+\hdots$, which satisfy the commutation relations in Eqs.~\ref{ladder 1} and \ref{ladder 2}.
The sole difference is that the individual terms, $H_a^{(n)}$, for the expansions in $\frac{1}{N\sqrt\beta}$ are some terms expressed in terms of both the bosonic field, $\varphi(z)$ and the fermionic field, $\psi(z)$.

\subsubsection{A Basis of Fields}

We proceed exactly as before, beginning with a definition of the basis we can use to express any local Hamiltonian constructed from these two fields, $\varphi(z)$ and $\psi(z)$, which are relevant to the problem.
These will now be labelled by a pair of partitions, $\bgam=\{\gamma_1,\gamma_2,\hdots\}$ and also $\bxi=\{\xi_1,\xi_2,\hdots\}$, one for the bosonic sector and one for the fermionic sector,
	\eq{T_{\bgam,\bxi} = \oint\cint{z}z^{|\bgam|+|\bxi|+\frac{l(\bxi)}{2}-1}
		\prod_{\gamma_i\in\bgam}i\partial^{\gamma_i}\varphi(z)
		\prod_{\xi_i\in\bxi}\partial^{\xi_i}\psi(z).}
There are a few things to note about this definition.
Firstly, $\bgam$, relating to the bosonic fields, is as before; it is a partition of positive integers which we order such that $\gamma_1\ge\gamma_2\ge\hdots$.
The new, fermionic partition, $\bxi$, is also a set of integers though it cannot contain any integer twice (due to fermionic exclusion) and it must have an even number of elements (due to parity symmetry).
We denote the number of elements by $l(\bxi)$ so, for example, $l(\{0,1\})=2$ and $l(\{0,1,2,3\})=4$.
Furthermore, it can also contain $0$ as an entry ($i\partial^0\varphi$ is excluded due to number conservation though no such symmetry prevents $\partial^0\psi$).
To further distinguish this partition from $\bgam$ we will also order it in reverse, with $\xi_1<\xi_2<\hdots$.
Finally, the product over the elements, $\xi_i$, must have a specific ordering, as permutations may bring about an overall sign change, so we define this product such that
	\eq{\prod_{\xi_i\in\bxi}\partial^{\xi_i}\psi = \partial^{\xi_1}\psi\partial^{\xi_2}\psi\hdots.}
Note that in both the bosonic and fermionic cases the empty set, $\emptyset$, refers to no contributions from that sector.
So for example $T_{\bgam,\emptyset}$ are all the purely bosonic terms which we found when considering the Laughlin state and include no fermionic fields.

Once again, there is a large degeneracy in this definition of basis elements.
For example,
	\alg{T_{\emptyset,02} & = \oint\cint{z} z^2\psi\partial^2\psi \nonumber\\
			& = - \oint\cint{z} \partial(z^2\psi)\partial\psi = -2T_{\emptyset,01}.}
Therefore, we must once again weed out these duplicate entries, and so we provide a particular choice of a linearly independent basis of terms in table \ref{MR terms table} for the first few scaling dimensions, where the scaling dimension of a given term is $d_{\bgam,\bxi} = |\bgam|+|\bxi|+\frac{l(\bxi)}{2}$.

\begin{table}[ht!]
	\begin{tabular}{| c | l |}
		\hline$d_{\bgam,\bxi}$ & Unique terms \\\hline\hline
		2 & $T_{11,\emptyset}, T_{\emptyset,01}$ \\
		3 & $T_{111,\emptyset}, T_{1,01}$ \\
		4 & $T_{22,\emptyset}, T_{1111,\emptyset}, T_{2,01}, T_{11,01}, T_{\emptyset,12}$ \\
		5 & $T_{221,\emptyset}, T_{11111,\emptyset}, T_{3,01}, T_{21,01}, T_{111,01},
				T_{1,12}$ \\
		$\vdots$ & \multicolumn{1}{|c|}{$\vdots$} \\ \hline
	\end{tabular}
	\caption{A table of some of the first few linearly independent terms from which $H$ can be constructed in the Moore-Read case.
	The number of possibilities grows much quicker with the scaling dimension than it did in the Laughlin case.
	}
	\label{MR terms table}
\end{table}

\subsubsection{The Leading Contributions}

We will now restrict the wide array of possible terms in table \ref{MR terms table} to those which satisfy the symmetries of our Hamiltonian.
Recall that to first order this entails searching for $H_a^{(0)}$ which commute with both $a_1$ and $a_{-1}$.
In the Laughlin case this was extremely restrictive.
However, in this Moore-Read case the constraint is somewhat less powerful due to the fact that the centre of mass mode commutes with terms of purely fermionic nature.
As such, there are more contributions than previously, as provided in table \ref{MR leading table}.

\begin{table}[ht!]
	\begin{tabular}{| c | c |}
		\hline$d_a$ & $H_a^{(0)}$ \\\hline\hline
		2 & $T_{\emptyset,01}$ \\
		3 & \\
		4 & $T_{22,\emptyset} \qquad T_{\emptyset,12}$ \\
		5 & $T_{3,01}+3T_{2,01}$ \\
		6 & $T_{33,\emptyset} \qquad T_{\emptyset,23}$ \\
		$\vdots$ & $\vdots$ \\ \hline
	\end{tabular}
	\caption{The leading contributions to the effective Hamiltonian in the Moore-Read case.
		These contributions occur at order $R^{1-d_a}$ and are mostly diagonal in the basis defined by Eq.~\ref{Moore-Read edge states}, and so primarily affect the dispersion.
		However, there is a scattering term at order $R^{-4}$ which couples the fermionic edge states with the bosonic modes.}
	\label{MR leading table}
\end{table}

Finding all the sub-leading contributions is then a case of applying the commutation relation in Eq.~\ref{ladder 2} to generate the ladder of terms, $H_a^{(1)}$, $H_a^{(2)}$ and so on.
As before, the situation is relatively simple, with $L_{-1}$ acting as a derivative on the fields within each $T_{\bgam,\bxi}$ on the right and $a_{-1}$ seeing only bosonic fields on the left.
Applying this procedure, one then finds that the three leading terms for Moore-Read hamiltonians have the form
	\eq{\boxed{H_{\emptyset,01} = T_{\emptyset,01} + \sum_{n=1}^\infty
			\frac{1}{(N\sqrt\beta)^n}\frac{(2n-1)!!}{(2n)!!}T_{2^n,01},} \label{H01 def}}
	\eq{\boxed{H_{22,\emptyset} = T_{22,\emptyset} + \sum_{n=1}^\infty
			\frac{8}{(N\sqrt\beta)^n}\frac{(2n+1)!!}{(2n+4)!!}T_{2^{n+2},\emptyset},} \label{H22 def}}
	\eq{\boxed{H_{\emptyset,12} = T_{\emptyset,12} + \sum_{n=1}^\infty
		\frac{1}{(N\sqrt\beta)^n}\frac{(2n+1)!!}{(2n)!!}T_{2^n,12}.} \label{H12 def}}
Therefore, each of the terms which, to leading order, simply change the dispersion of the fermionic modes (i.e, those of the form $H_{\emptyset,\bxi}$) necessarily have sub-leading contributions which couple the bosonic and fermionic excitations, suppressed by a factor of $\frac{1}{N\sqrt\beta}$.

\section{Numerics}
\label{numerics}

We once again stress that the claim of locality which was so instrumental in calculating these allowed contributions to our Hamiltonian was a conjecture based primarily on the understanding of the Laughlin state as a plasma in its screening phase (for $\frac{1}{\nu}\lesssim 65$).
As such, we present thorough numerical evidence to further motivate this claim by comparing the results of these local effective Hamiltonians, which we hope will provide a very accurate description of the low-energy physics, with exact numerical results\cite{theCode}.

To asses our effective Hamiltonians we will take $H$ to be simply a sum of the first few, least irrelevant terms from the field theoretic considerations we have made, including their coupling coefficients, $h_a$, which we cannot fix by symmetry alone.
These Hamiltonians are then simple to diagonalise, being phrased in terms of second quantised operators.
We then perform the exact numerics by generating the full edge states in a basis of single-particle states, the monomial basis, by using their expression in terms of Jack polynomials\cite{bernevig2008model, bernevig2008generalized, bernevig2008properties}.
We then exactly diagonalise the interaction within reduced subspaces of edge states at fixed angular momentum to find the eigenstates and eigenvalues, and compare to the effective Hamiltonians, $H(h_a)$, by fitting the coupling coefficients to the data.

\subsection{Confinement}

\begin{figure}[ht!]
	\centering
	\includegraphics[width=\linewidth]{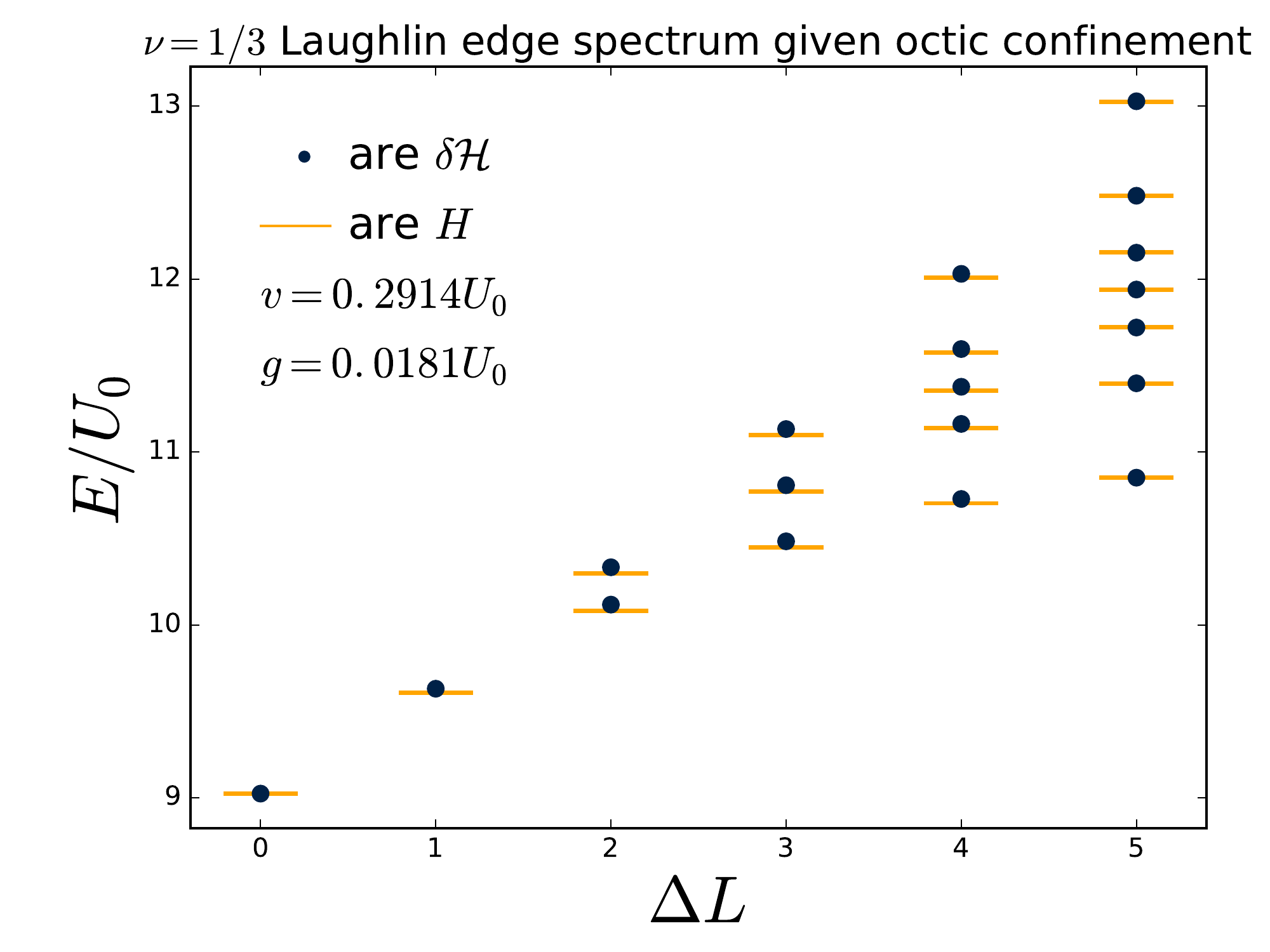}
	\caption{A comparison of our effective Hamiltonian with fit parameters $g$ and $v$ with the exact spectrum for an $N=10$ Laughlin state at $\nu=1/3$ confined purely by a weak octic confinement $U = U_0\sum_i\lr{r_i/R}^8$ (i.e, there is no additional quadratic confinement).
	}
	\label{confinement spectrum}
\end{figure}

We will initially consider the simpler case of confinement where the perturbation is simply a single-body term in the Hamiltonian.
Recall then that we cannot use translational invariance to significantly constrain the final form of the effective Hamiltonian so we simply take the first few least irrelevant contributions,
	\alg{H = \oint\cint{z}&\lr{\frac{v}{2}z\lr{i\partial\varphi(z)}^2 + 
				g z^2\lr{i\partial\varphi(z)}^3} \nonumber\\&
		+ \order{R^{-3}}.}
We will use this to consider a relatively steep edge confinement of $U=U_0\lr{r/R}^8$ and fit the coefficients $v$ and $g$ by simply matching the exact and effective spectra.
We find the results in Fig.~\ref{confinement spectrum} for the $\nu=1/3$ Laughlin state containing $N=10$ particles, and find a very good match. 

\begin{figure}[ht!]
	\centering
	\includegraphics[width=\linewidth]{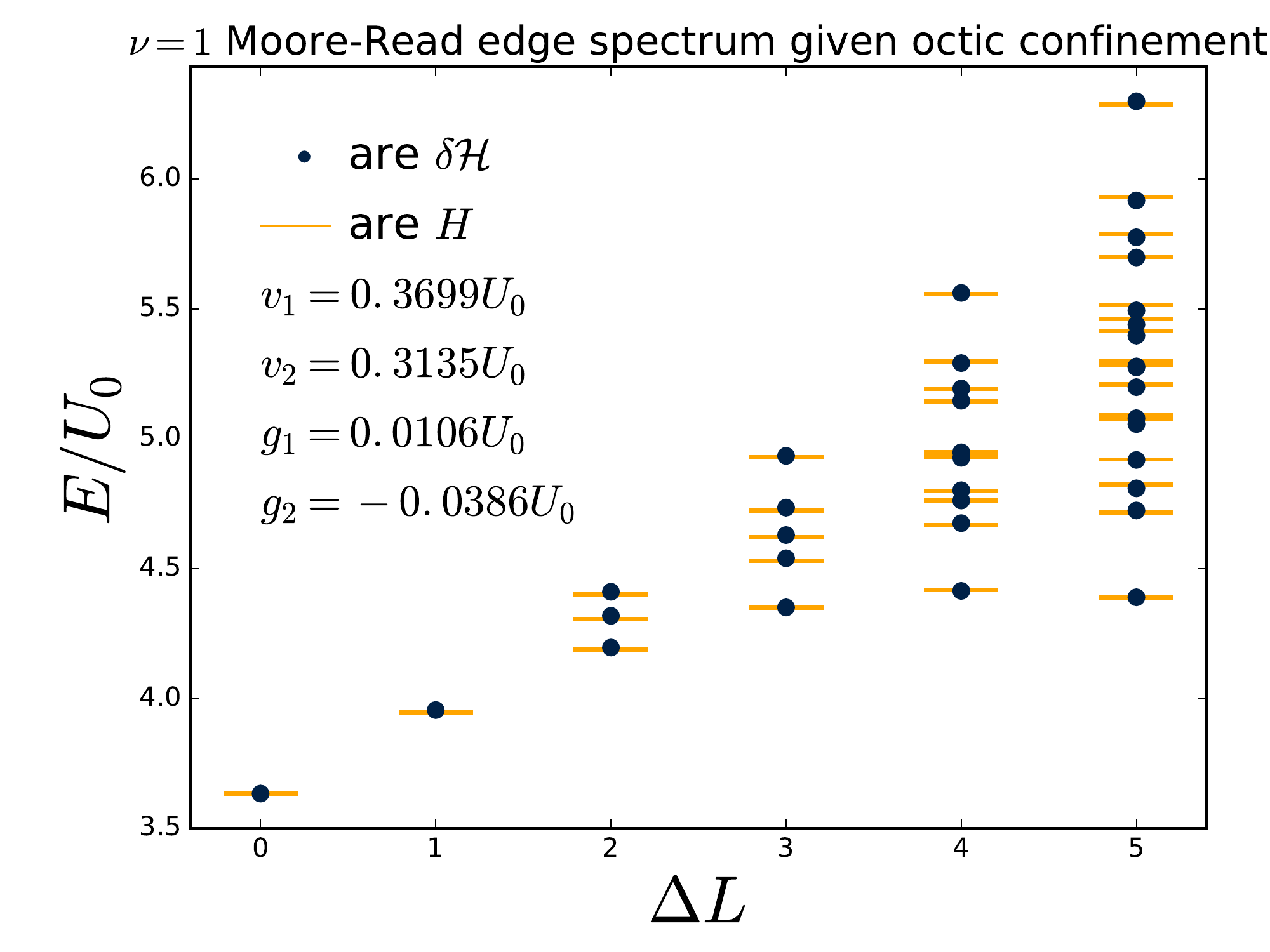}
	\caption{A comparison of our effective Hamiltonian with fit parameters $v_1, v_2, g_1$ and $g_2$ with the exact spectrum for a $\nu=1$ Moore-Read state confined by a weak octic confinement $U = U_0\sum_i\lr{r_i/R}^8$ when $N=14$.
	Once again, the quadratic confinement was set to zero for these results.}
	\label{MR confinement}
\end{figure}

We also consider the Moore-Read case.
Once again, without translational invariance the resulting effective Hamiltonian cannot be significantly simplified, which leaves us with the following least irrelevant terms,
	\alg{H = \oint\cint{z}&\left(-\frac{v_1}{2}z\lr{\psi\partial\psi(z)} + \frac{v_2}{2}z\lr{i\partial\varphi(z)}^2\right.\nonumber\\&
			\left. + g_1 z^2\lr{i\partial\varphi(z)}^3 + g_2 z^2\lr{i\partial\varphi\psi\partial\psi(z)}\right) \nonumber\\&
		+ \order{R^{-3}}.}
Once again, we check this against exact numerical results for a Moore-Read droplet confined by radial potential of the form $U_0\lr{r/R}^8$.
The results are given in Fig.~\ref{MR confinement} and show good agreement once again.

\subsection{Interacting Laughlin}

For the Laughlin state perturbed by nontrivial interactions we shall employ a simple two-parameter effective Hamiltonian, including only the two least irrelevant contributions,
	\eq{H = h_{22}H_{22} + h_{33}H_{33}. \label{Laughlin effective H}}
This should replicate the matrix elements of the Hamiltonian as calculated numerically up to corrections of order $R^{-7}$.

We then note that, given this form for the effective Hamiltonian, certain matrix elements will constrain the coefficients $h_{22}$ and $h_{33}$ exactly.
Specifically, consider the element $\bra{2}H\ket{2}$ where $\ket{2}=a_{-2}\ket{0}$ as defined in Eq.~\ref{Laughlin edge states}.
The only non-zero contribution to this comes from $T_{22}$ contained in $H_{22}$.
Not only is this true for the truncated expansion, but we expect it to hold true at every order.
Furthermore, the only local terms which we expect can contribute to the matrix element $\bra{3}H\ket{3}$, even in an infinite-order expansion, are $T_{22}$ and $T_{33}$.
Therefore, these two matrix elements should determine $h_{22}$ and $h_{33}$ exactly, and these are what we use to fit these coefficients.

Note that to find the effective Hamiltonian $H$ from the data requires one to also calculate the overlaps of the wavefunctions.
To see this, recall that the states in the quantum Hall language are not orthogonal, i.e,
	\eq{\bbrakket{\Psi_{\bra{v}}}{\Psi_{\bra{w}}} = \bra{w}G_N\ket{v}.}
The quadratic form $G_N$ was discussed in length in Refs.~\onlinecite{dubail2012edge,fern2018structure}. Furthermore,
	\eq{\bbra{\Psi_{\bra{v}}}\delta\ham\kket{\Psi_{\bra{w}}} = \bra{w}HG_N\ket{v}.}
Therefore, we can calculate the matrices $G_N$ and $(HG_N)$, and therefore we find the effective Hamiltonian by
	\eq{H = (HG_N)G_N^{-1}.}

\subsubsection{Coefficient Scaling}

\begin{figure}[ht!]
	\centering
	\includegraphics[width=\linewidth]{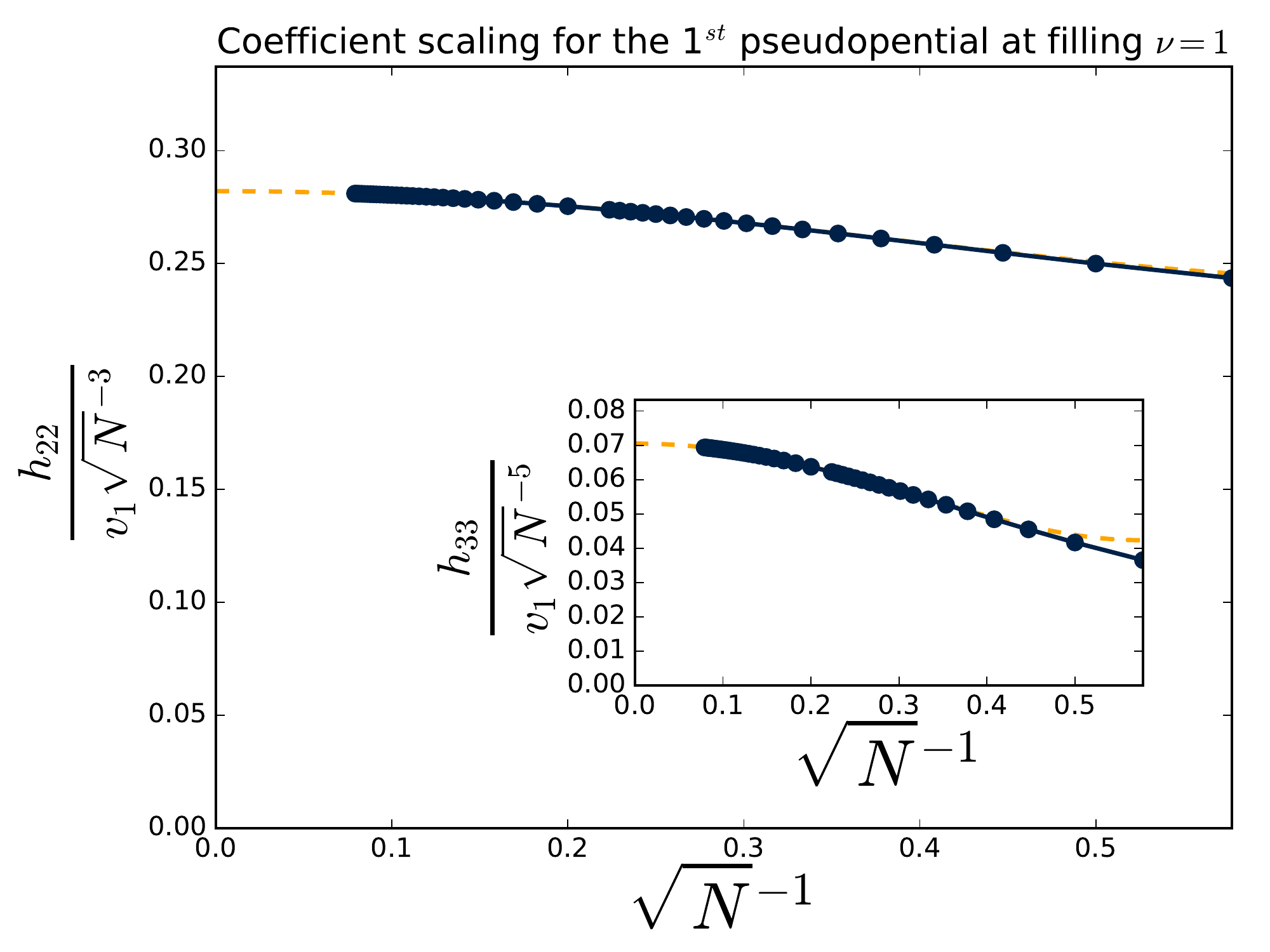}
	\caption{Perturbing the integer quantum Hall effect at $\nu=1$ with a first pseudopotential, $V_1$, we find the above scaling of the coefficients $h_{22}$ and $h_{33}$ in the effective Hamiltonian.
	In both cases the coefficients appear to obey the scaling hypothesis well, varying as $h_a\sim \sqrt N^{1-d_a}$ where $d_{22}=4$ and $d_{33}=6$.
}
	\label{integer coeffcients}
\end{figure}

\begin{figure}[ht!]
	\centering
	\includegraphics[width=\linewidth]{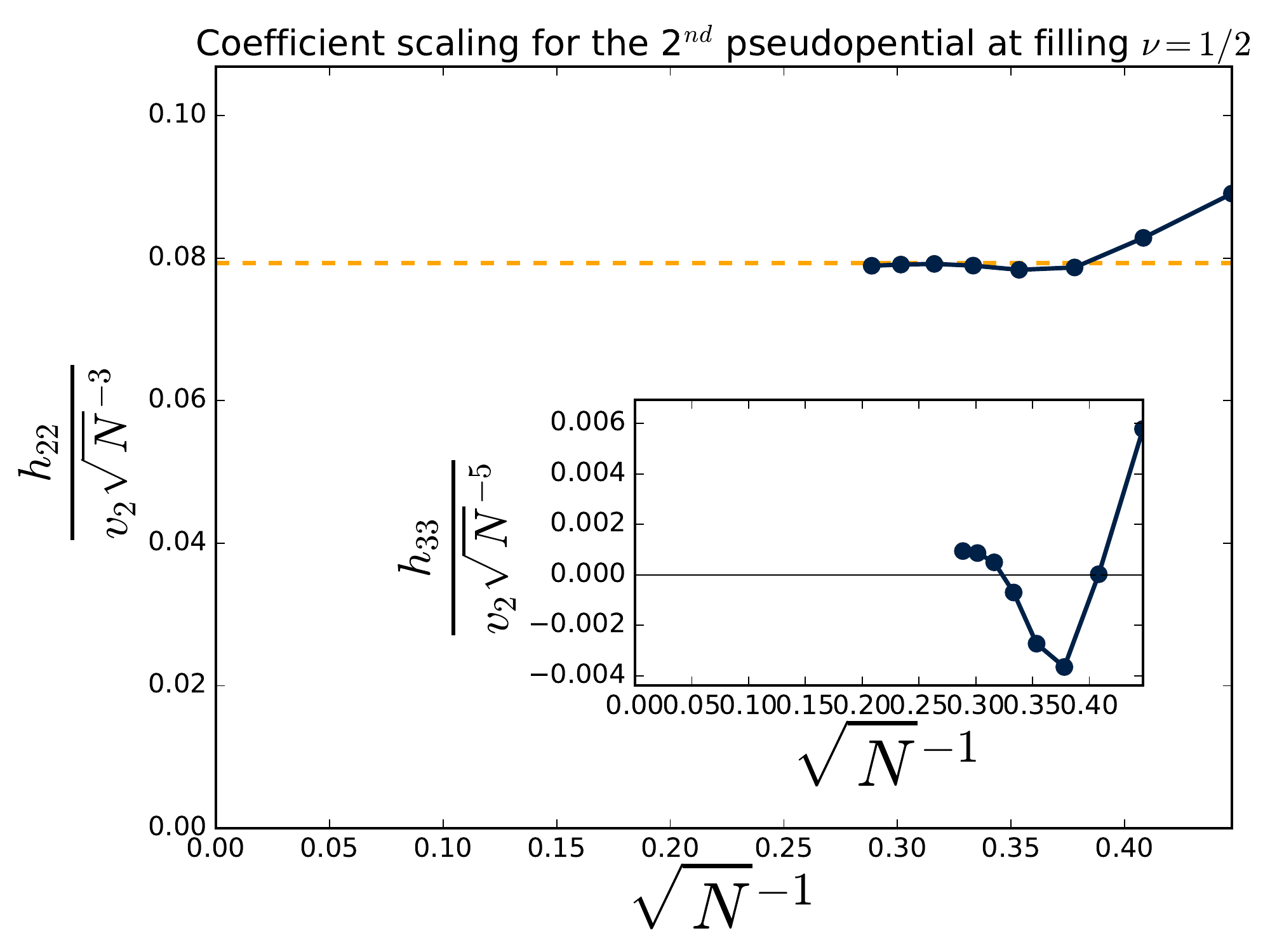}
	\caption{We perturb the Laughlin wavefunction at $\nu=1/2$ by the second pseudopotential.
		Plotted here is the value we fit for $h_{22}$ (which we expect to scale as $v_2\sqrt N^{-3}$) in the effective Hamiltonian as $N$ is varied, which appears to converge very quickly to some $\tr{constant}/\sqrt{N}^3$.
		Unfortunately, $h_{33}$, which is expected to vary as $\sqrt{N}^{-5}$, does not converge for this range of system sizes but does appear to decay at least as quickly as $\sqrt{N}^{-5}$, if not faster, as required by the scaling hypothesis.
}
	\label{h22 pseudo2}
\end{figure}

Upon fitting the coefficients it is important to check that the scaling arguments made previously hold.
These claimed that the coefficients $h_a$ should scale as $R^{1-d_a}$.
Then, given that the radius scales as the square root of particle number, $N$, means
	\eq{h_a \sim \sqrt{N}^{1-d_a}.}
To check that this scaling is borne out in the data, we fit the parameters for a variety of system sizes for fillings $\nu=1$ and $\nu=1/2$ and plot the results in Figs.~\ref{integer coeffcients} and \ref{h22 pseudo2}.
The exact results will depend upon the interaction and filling and in these examples we take the interaction to be the first Haldane pseudopotential for $\nu=1$ and the second Haldane pseudopotential at $\nu=1/2$.
We see that the results appear to be consistent with the scaling hypothesis, with $h_{22}$ varying in the large-$N$ limit as $h_{22}\sim \sqrt{N}^{-3}$, exactly as expected.
$h_{33}$ is less clear.
For the integer case it also appears to vary as expected, $h_{33}\sim\sqrt{N}^{-5}$, but at $\nu=1/2$ it appears that sub-leading corrections to this coefficient are large enough that the value does not converge for the range of $N$ we can reach with exact methods.
Nevertheless it does not appear to fall off slower than $\sqrt{N}^{-5}$, so this also appears consistent with scaling arguments.

\subsubsection{Effective Hamiltonian Spectra}

\begin{figure}[ht!]
	\centering
	\includegraphics[width=\linewidth]{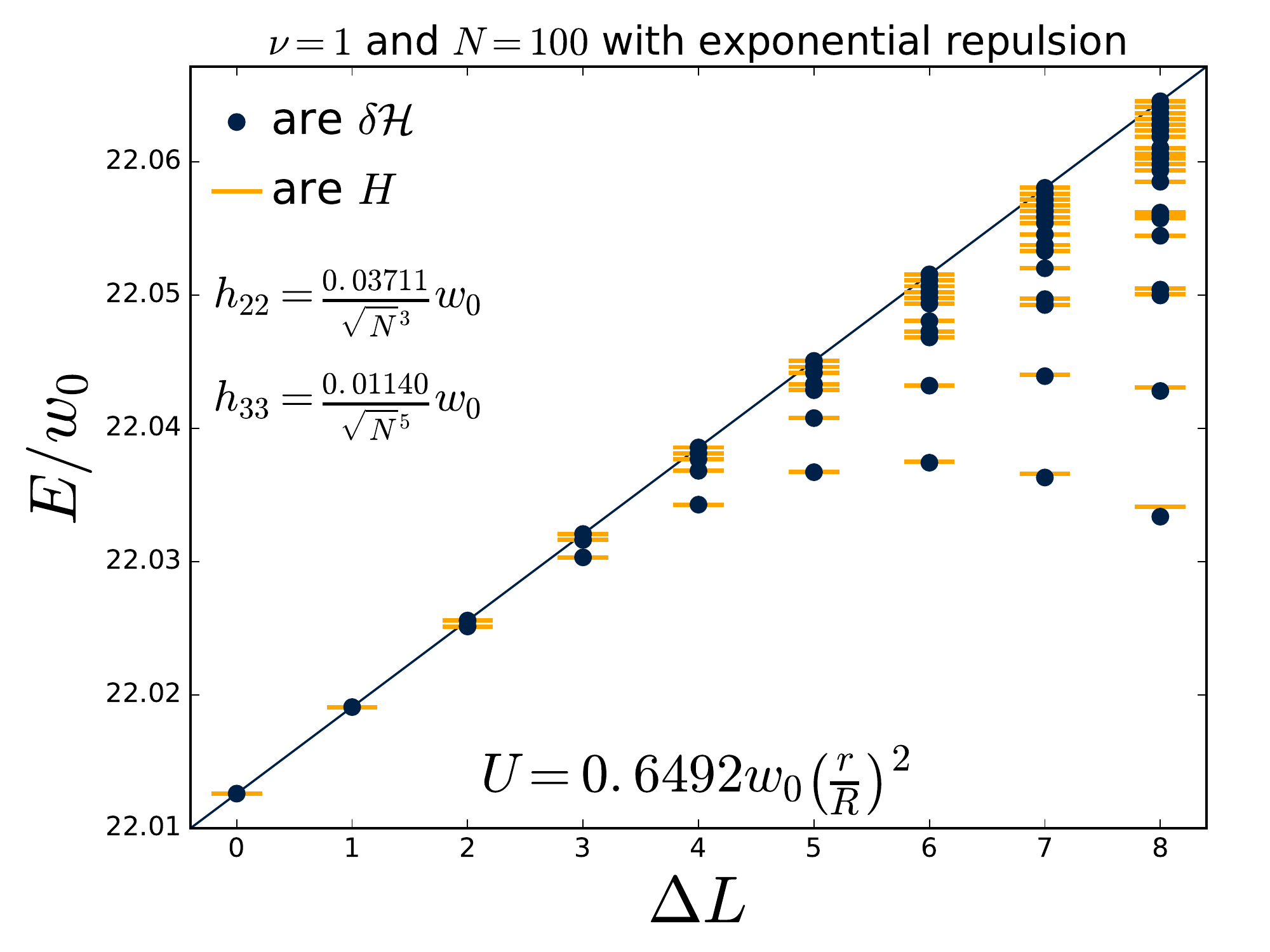}
	\caption{A comparison of our 2-parameter effective Hamiltonian with the exact edge spectrum for a $\nu=1$ quantum Hall state perturbed by exponential repulsive interactions in Eq.~\ref{exponential repulsion} in the limit where $w_0$ is small.
	Recall that, because we expect the only terms in $H$ which contribute to the matrix elements we use to fit $h_{22}$ and $h_{33}$ are $H_{22}$ and $H_{33}$, we expect that the fits to these coefficients are exact for these systems at these particle numbers.
	We find that the subsequent agreement is extremely good, capturing the distinct characteristics of the spectrum and matching most points extremely closely.
	}
	\label{nu1_exp}
\end{figure}

\begin{figure}[ht!]
	\centering
	\includegraphics[width=\linewidth]{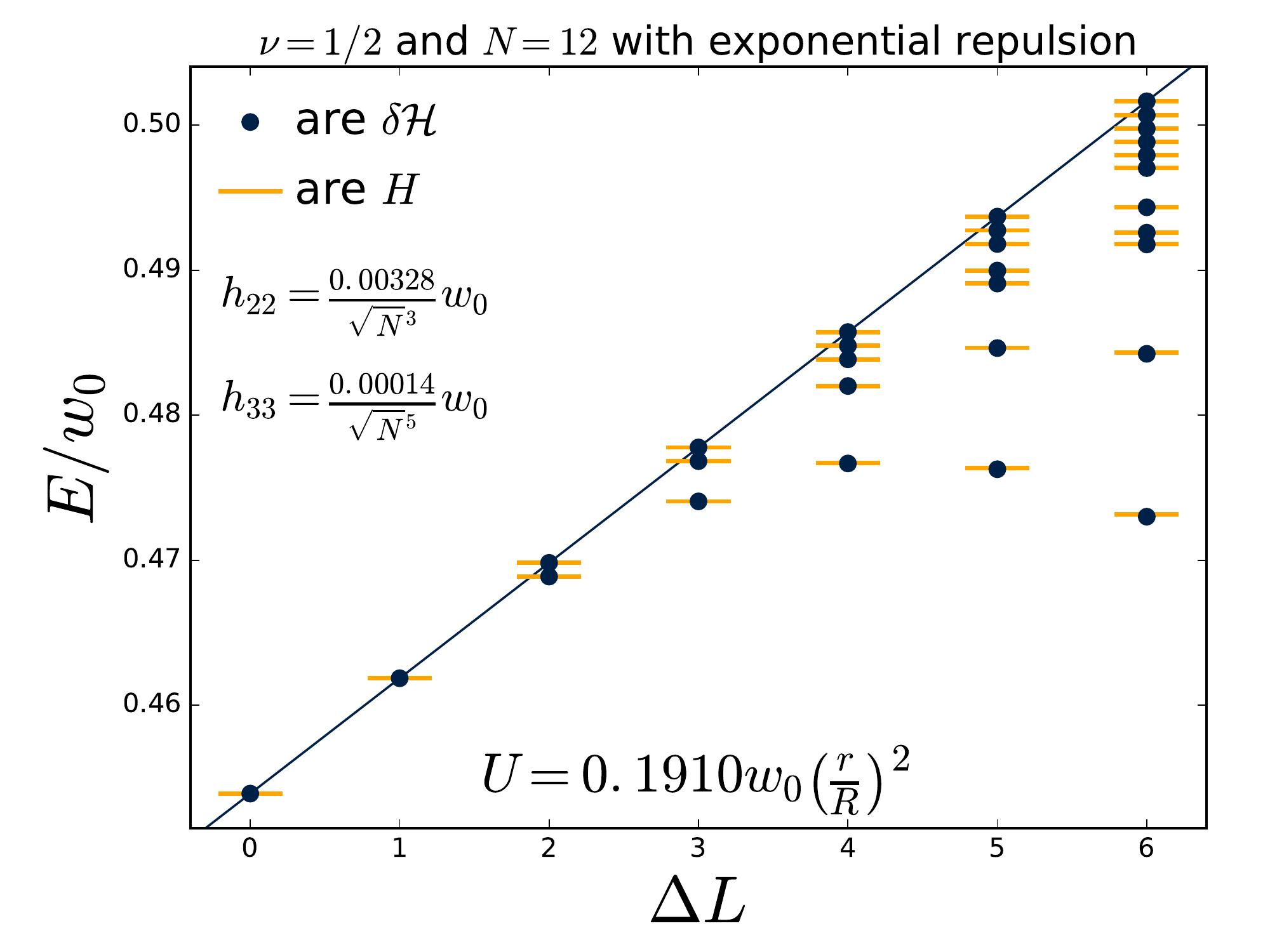}
	\caption{A comparison of numerical results for a $\nu=1/2$ Laughlin state perturbed by exponential interactions with our 2-parameter effective Hamiltonians.
		As with the integer quantum Hall case, the agreement is excellent.}
	\label{nu2_exp}
\end{figure}

\begin{figure}[ht!]
	\centering
	\includegraphics[width=\linewidth]{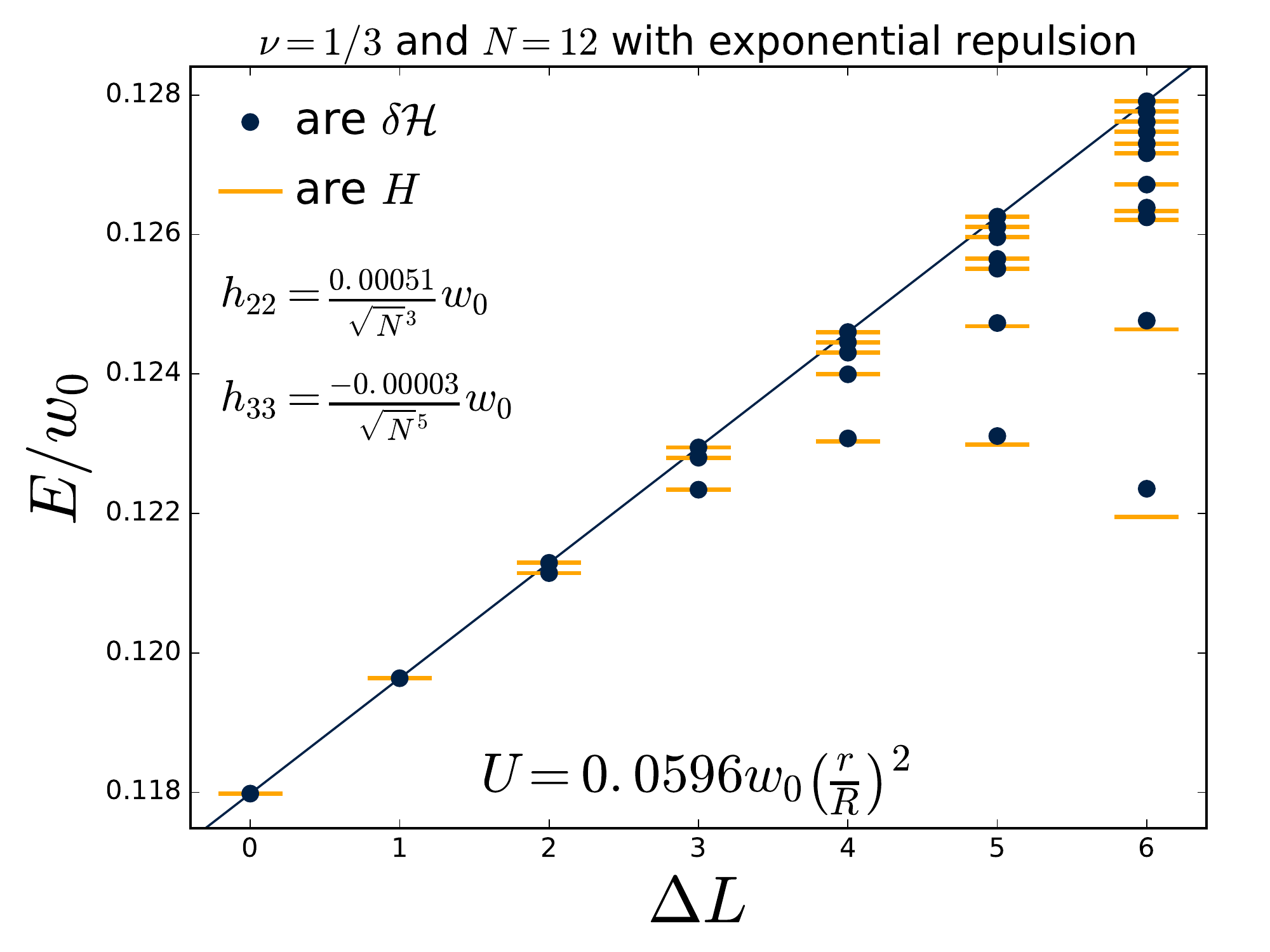}
	\caption{A final comparison of numerical results with our effective description for the $\nu=1/3$ Laughlin state, which once again shows excellent agreement.}
	\label{nu3_exp}
\end{figure}

Perhaps the most crucial check that our effective Hamiltonians describe the true behaviour of quantum Hall edges is that they faithfully reproduce the spectrum of edge states.
Therefore, we take the Hamiltonian in Eq.~\ref{Laughlin effective H} and fit values for $h_{22}$ and $h_{33}$ based on the method described above.
We then plot the agreement for the case of exponential repulsion between particles,
	\eq{V = w_0\sum_{i<j}\exp\lr{-\left|\frac{z_i-z_j}{2\ell_B^2}\right|^2},\label{exponential repulsion}}
at a variety of filling fractions in figures \ref{nu1_exp}, \ref{nu2_exp} and \ref{nu3_exp}.
In each case we find that this two-parameter fit is very good.

Note that in these plots the linear slope is a free parameter of the parent Hamiltonian derived from the confinement, which we always take as quadratic.
Therefore, in each plot we add an arbitrary harmonic potential (specified in each figure) which produces the accompanying linear slope.
The repulsive interactions then cause the energies to decrease with respect to this linear edge as angular momentum is increased.
This is because these excited states allow the particles to avoid each other more successfully, increasing their average separations.
The states which then lie exactly on the Luttinger liquid line are those edge states which correspond simply to translations of the original circular droplet as a whole, i.e, the states $a_{-1}^{\Delta L}\ket{0}$.

\subsubsection{Constraints on non-local terms}

Throughout the preceding text we have assumed that the only terms which contribute to $H$ are local.
However, this is a conjecture based on the Laughlin state being in a screening phase for $\beta\lesssim65$, which makes correlations short range.
However, for sufficiently long-range interactions we expect non-local terms to also contribute.
One such example is the harmonic interaction, for which we can find the effective Hamiltonian exactly as presented in Appendix \ref{Harmonic Interactions} and which contains non-local contributions.

However, this is a special case of an interaction which actually grows with separation.
In a more general setting, one of the most well-known non-local terms we might expect to contribute is the Benjamin-Ono term\cite{bettelheim2006nonlinear, wiegmann2012nonlinear, abanov2005quantum}, which has the form
	\alg{T_\tr{B-O} & = \underset{|z|>|w|}{\oint\cint{z}\oint\cint{w}}\frac{zw}{(z-w)^2}:i\partial\varphi(z)i\partial\varphi(w):
				\nonumber \\&
				= \sum_{n>0}na_{-n}a_n}
and has the lowest scaling dimension possible for such a double-integral term.
Therefore, to ascertain the likelihood that non-local terms might appear we insert it into the Hamiltonian and attempt to fit its coefficient.
We will then analyse the scaling of the resulting coefficient.

Therefore, we take the ultra-simple Hamiltonian
	\eq{H = g_\tr{B-O}\lr{T_\tr{B-O}-\frac{1}{2}T_{11}} + g_{22}T_{22} + \hdots \label{non-local_H}}
and fit these two coefficients.
Note that the inclusion of the $T_{11}$ here ensures that the overall Benjamin-Ono term which we have inserted here obeys the leading condition of translational symmetry, i.e,
	\eq{[a_1,H_a^{(0)}]=[a_{-1},H_a^{(0)}] = 0.}
The terms that we throw away in Eq.~\ref{non-local_H} are then either off-diagonal (terms which are generated by this Benjamin-Ono term to satisfy translational invariance at all orders) or of order $N^{-5/2}$.
We then fit the coefficients $g_\tr{B-O}$ and $g_{22}$ with the first two non-trivial diagonal elements, $\bra{2}H\ket{2}$ and $\bra{3}H\ket{3}$.

\begin{figure}[ht!]
	\centering
	\includegraphics[scale=0.4]{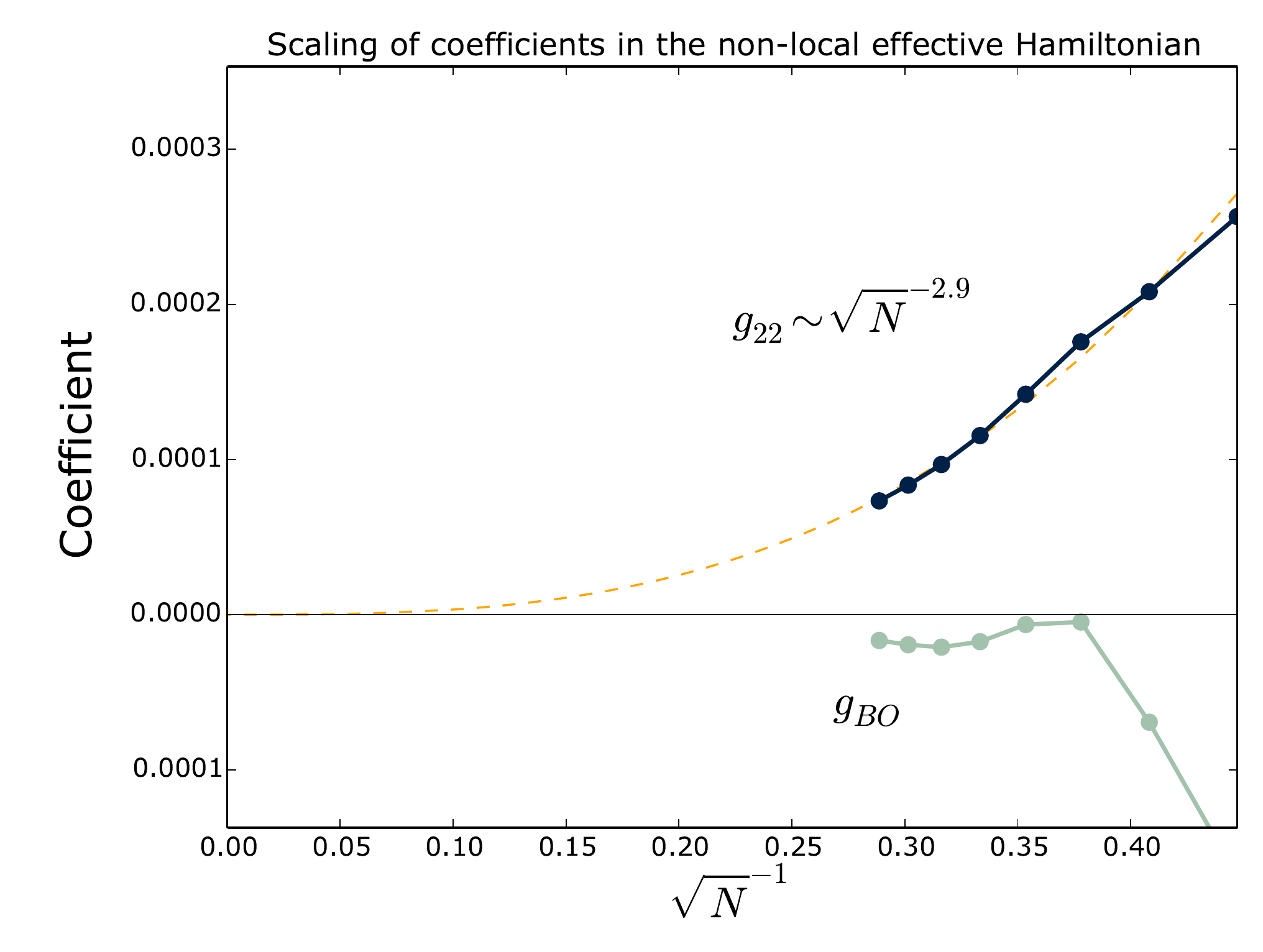}
	\caption{We assume for the moment that the Hamiltonian might contain non-local terms and fit them taking the form in Eq.~\ref{non-local_H} for $H$ and taking exponential interactions between the particles at filling $\nu=1/2$.
	We see that the Benjamin-Ono coefficient is consistently smaller than $g_{22}$ despite scaling arguments suggesting that it should be $\sim \sqrt N$ times larger.
	As such, any contribution this term makes to the dynamics are much smaller than expected.
	Clearly this does not prevent any non-local term being present at any order in the expansion but it does provide evidence against this most likely contribution.}
	\label{BO_Hamiltonian}
\end{figure}

The fits for the coefficients are plotted in Fig.~\ref{BO_Hamiltonian} for the case of an exponential interaction at $\nu=1/2$.
We see that the scaling of $g_{22}$ is very close to the expected $\sqrt{N}^{-3}$ whereas the Benjamin-Ono coefficient does not scale in a manner which can even be approximated by a power law.
Nevertheless, we see that it is always much smaller than $g_{22}$ despite the fact that it is expected to be roughly $\sqrt N$ times larger.
Therefore, it is unlikely that the Benjamin-Ono term contributes to our effective Hamiltonian, or at least its effects are much smaller than expected by simple scaling arguments.

\subsubsection{Couplings for Pseudopotentials}

We now look at what the values for the couplings, $h_{22}$ and $h_{33}$, look like for the first few Haldane pseudopotentials.
In theory, each pseudopotential has associated with it an effective Hamiltonian,
	\alg{V_k\Psi_{\bra{\blam}} = \Psi_{\bra{\blam}H_k} \label{H definition},}
and therefore, given that any interaction can be formed from a sum of pseudopotentials, i.e, $V=\sum_kv_kV_k$, the effective Hamiltonian of such an interaction is simply
	\eq{H = \sum_kv_kH_k. \label{pseudopotential effectives}}
Recall that the parent Hamiltonian at $\nu=1/m$ is constructed from all the pseudopotential $V_k$ with $k<m$ and so these annihilate all our wavefunctions at these filling fractions.
Therefore, knowledge of the coupling coefficients within the $H_k$ of Eq.~\ref{pseudopotential effectives} for $k\ge m$ is all we require to be able to build the effective Hamiltonian for any interaction.

Unfortunately, as we have already seen in Fig.~\ref{h22 pseudo2} for example, fitting some of these coefficients can be difficult or simply unreliable for the system sizes we are able to compute exactly.
Therefore, we fit only the leading order contribution, $h_{22}$ for each and give the values in table \ref{g22_table} for the pseudopotentials contributing to the interactions at $\nu=1/2$.

\begin{table}[ht!]
	\centering
	\begin{tabular}{|c||c|}
		\hline Pseudopotential & $h_{22}^{\nu=1/2}$ \\ \hline\hline
$V_2$ & $0.079\pm0.007$ \\ \hline
$V_4$ & $0.074\pm0.020$ \\ \hline
$V_6$ & $0.087\pm0.036$ \\ \hline
	\end{tabular}

	\caption{The leading coupling coefficient for the effective Hamiltonian of various pseudopotentials at fractional fillings $\nu=1/2$.
		For the case at half filling the convergence to a constant appears robust and the errors are very small.
		Similar extrapolations could not be made reliably for the case at $\nu=1/3$ given the inferior convergence in this case.}
	\label{g22_table}
\end{table}

However, whilst our data for fractional fillings remains too small to form a reliable conclusion about the sub-leading corrections to the effective Hamiltonian, we can perform this analysis extremely reliably for $\nu=1$, where we consider $N$ as high as 160.
Thus, we fit the effective Hamiltonian to 5$^\tr{th}$ order, expanding the leading coupling coefficient as
	\eq{h_{22} = \frac{h_{22}^{(3)}}{\sqrt N^3}+\frac{h_{22}^{(4)}}{\sqrt N^4}+\frac{h_{22}^{(5)}}{\sqrt N^5}+\order{\sqrt N^{-6}}}
and considering only the leading order for $h_{33}=h_{33}^{(5)}/\sqrt N^5+\order{\sqrt N^{-6}}$.
Each of these coefficients is shown in table \ref{nu1_coeffs}.

\begin{table*}[t!]
	\centering
	\begin{tabular}{|c||c|c|c|c|}
		\hline Pseudopotential & $h_{22}^{(3)}$ & $h_{22}^{(4)}$ & $h_{22}^{(5)}$ & $h_{33}^{(5)}$ \\ \hline\hline
$V_1$ & $0.282\pm0.001$ & $0.000\pm0.002$ & $-0.179\pm0.016$ & $0.071\pm0.001$ \\ \hline
$V_3$ & $0.423\pm0.001$ & $0.001\pm0.017$ & $-0.633\pm0.187$ & $0.247\pm0.001$ \\ \hline
$V_5$ & $0.529\pm0.001$ & $0.004\pm0.014$ & $-1.268\pm0.156$ & $0.485\pm0.001$ \\ \hline
$V_7$ & $0.617\pm0.001$ & $0.010\pm0.010$ & $-2.088\pm0.135$ & $0.771\pm0.005$ \\ \hline
$V_9$ & $0.694\pm0.006$ & $0.014\pm0.278$ & $-2.982\pm5.472$ & $1.103\pm0.189$ \\ \hline
	\end{tabular}
	\caption{
		The coupling coefficients for the first few contributing pseudopotentials at filling $\nu=1$.
			We note that the higher-order coefficients are larger relative to the lower order coefficients for the higher pseudopotentials (which are the pseudopotentials that are more important for less local interactions), i.e, $\left|h_{22}^{(3)}/h_{22}^{(5)}\right|$ is smaller for $V_k$ where $k$ is larger.
			Furthermore, we note that the coefficient of $h_{22}$ at order $\sqrt N^{-4}$ is very small relative to the coefficients at fractional powers of $N$.
			This is in good agreement with our calculation for the asymptotic behaviour of the exponential potential shown in a future publication \cite{bondesanFUTUREinteger}, which predicts vanishing coefficients at even order (i.e, at order $\sqrt N^{-2n}$).}
	\label{nu1_coeffs}
\end{table*}

\subsection{Interacting Moore-Read}

The spectra for the Moore-Read state are more complex than those for the Laughlin state.
For example, consider the subspace of states with $\Delta L=5$ units of angular momentum added with respect to the ground state.
In the Laughlin case, this is a 7-dimensional subspace but in the Moore-Read case there are 16 states.
As such, the matrix, $\delta\ham$, which our effective Hamiltonian, $H$, must reproduce includes over four times as many matrix elements.
Nevertheless, as we shall see, an effective Hamiltonian containing only three terms (and thus three fit parameters) can still provide an extremely good description of the resulting behaviour.
Thus, in the resulting discussion we will take the effective Hamiltonian to include the three least irrelevant terms
	\eq{H = h_{\emptyset,01}H_{\emptyset,01} + h_{22,\emptyset}H_{22,\emptyset}
			+ h_{\emptyset,12}H_{\emptyset,12}. \label{MR Hamiltonian fit}}
Note that to leading order, each term of this Hamiltonian is purely fermionic or bosonic, and hence the modes appear to decouple. However, one should recall that to first order each of these fermionic terms (as given in Eqs.~\ref{H01 def} and \ref{H12 def}) will couple the bosonic and fermionic edge channels.

\begin{figure}[ht!]
	\centering
	\includegraphics[width=\linewidth]{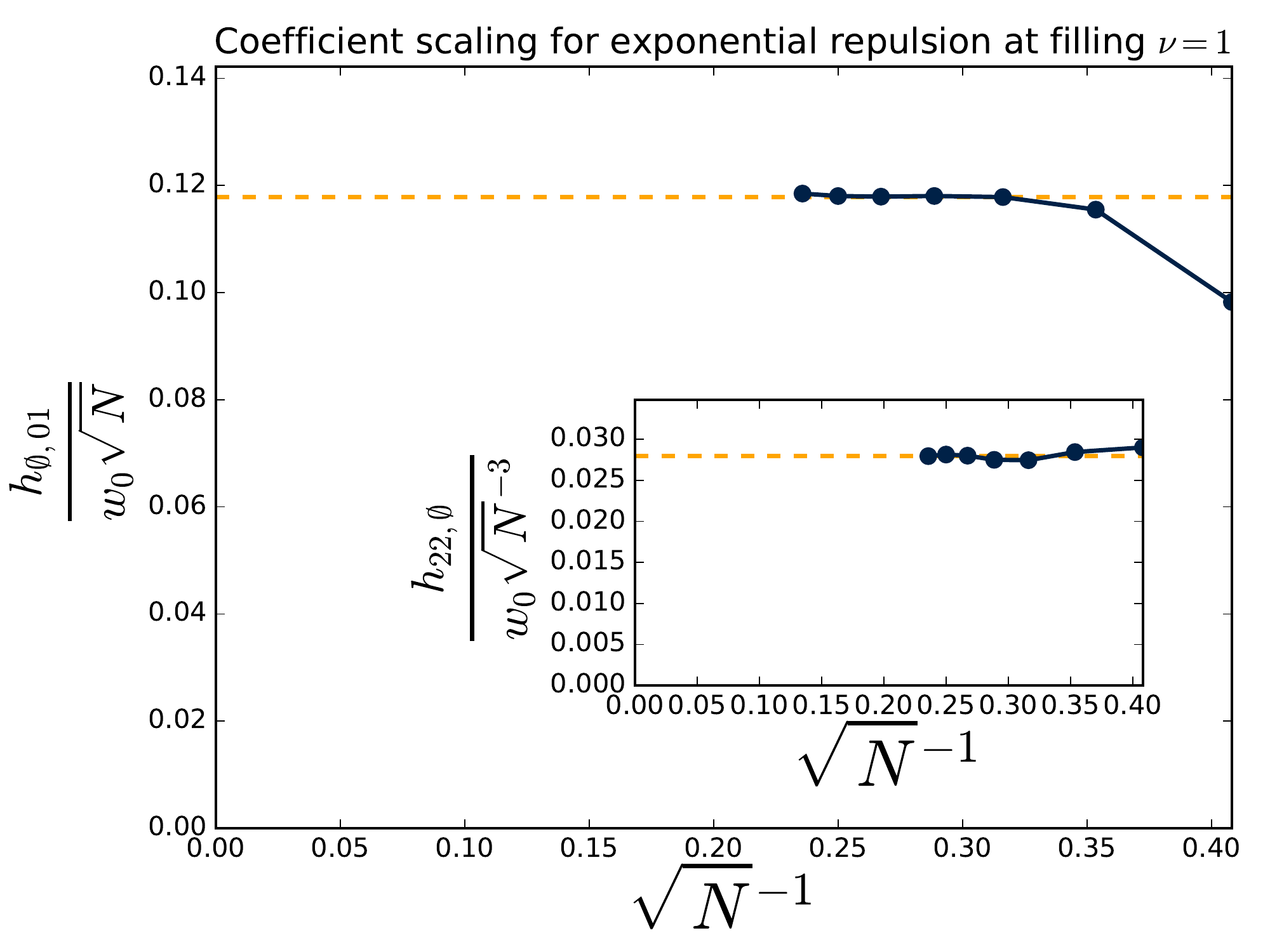}
	\caption{We consider the Moore-Read state at $\nu=1$ perturbed by an exponential repulsive interaction, Eq.~\ref{exponential repulsion}, and fit the coupling coefficients, $h_{\emptyset,01}, h_{22,\emptyset}$ and $h_{\emptyset,12}$ at a variety of system sizes, $N$.
	We then plot the variation of $h_{\emptyset,01}$ and $h_{22,\emptyset}$ with $N$, which are expected to vary as $\sqrt N^{-1}$ and $\sqrt N^{-3}$ respectively based on scaling arguments.
	The convergence to this scaling behaviour is quite good for both $h_{\emptyset,01}$ and $h_{22,\emptyset}$.
	However, the convergence is unclear for the coefficient $h_{\emptyset,12}$, whose scaling behaviour we do not show here, as we cannot reach large enough $N$ with these exact methods for the value to converge.
	Nevertheless, the scaling hypothesis does not appear to be incompatible with any of this data.
}
	\label{MR scaling}
\end{figure}

As in the Laughlin case, we will fit the coupling coefficients in Eq.~\ref{MR Hamiltonian fit} by comparison with particular matrix elements.
Specifically $\bra{2;\emptyset}H\ket{2;\emptyset}$ will fit the coefficient $h_{22,\emptyset}$.
Additionally, we use $\bra{\emptyset;\frac{3}{2}\frac{1}{2}}H\ket{\emptyset;\frac{3}{2}\frac{1}{2}}$ and $\bra{\emptyset;\frac{5}{2}\frac{1}{2}}H\ket{\emptyset;\frac{5}{2}\frac{1}{2}}$ together to fit the coefficients $h_{\emptyset,01}$ and $h_{\emptyset,12}$.
We present a scaling analysis of the resulting fit coefficients for the bosonic $\nu=1$ Moore-Read state perturbed by exponential interactions in Fig.~\ref{MR scaling}.
As these plots show, the scaling hypothesis appears to work well even for these relatively small system sizes.
The convergence for the least irrelevant contribution, $h_{\emptyset,01}$ is very good and is shown to vary as $\sqrt N^{-1}$ as expected by scaling arguments.
Less clear are the forms of scaling for $h_{22,\emptyset}$ and $h_{\emptyset,12}$ which do not converge so convincingly over the system sizes we are able to access though still appear to fall off no faster than the $\sqrt{N}^{-3}$ required.

\subsubsection{Effective Hamiltonian Spectra}

\begin{figure}[ht!]
	\centering
	\includegraphics[width=\linewidth]{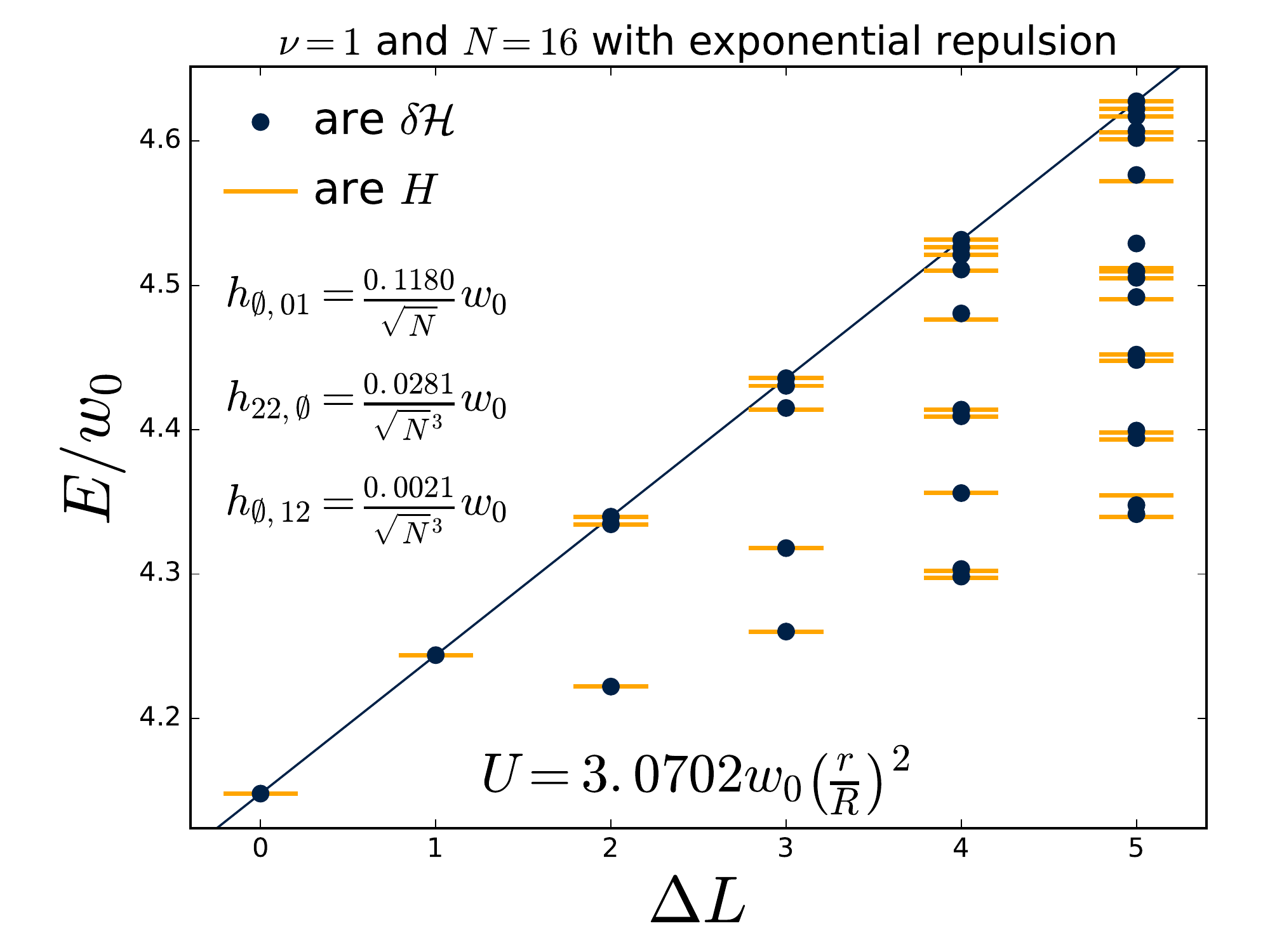}
	\caption{We show the spectrum for the Moore-Read state containing $N=16$ particles at filling $\nu=1$ perturbed by an exponential repulsive interaction, Eq.~\ref{exponential repulsion}.
	The three coupling coefficients in the effective Hamiltonian, $H$, are fit using the process described above and provide a very accurate fit to the numerically calculated spectrum (the dots corresponding to the spectrum of $\delta\ham$ whilst the orange lines are the spectra of $H$).
	As in the Laughlin case, the gradient of this linear Luttinger liquid slope (the blue line corresponding to an unperturbed droplet) is a free parameter of the parent Hamiltonian arising from the assumption of quadratic confinement.
}
	\label{MR spectrum}
\end{figure}

One again, the most crucial check of our effective theory is that they are able to reproduce the spectra of the corresponding systems.
As such, we calculate the spectrum numerically for Moore-Read states containing $N=16$ particles at filling $\nu=1$.
The data when the interactions we perturb with are exponentially repulsive (i.e, the same as in Eq.~\ref{exponential repulsion}) is shown in Fig.~\ref{MR spectrum} alongside a comparison to the effective Hamiltonian, $H(h_{\emptyset,01},h_{22,\emptyset},h_{\emptyset,12})$, where these coupling coefficients are fit using the procedure described above.

This comparison shows very good agreement between the exact numerical data and our low energy effective theories.
Notably, the renormalisation of the velocity of the fermionic modes is indeed borne out by the data, with the lowest energy modes in Fig.~\ref{MR spectrum} corresponding to cases where all of the angular momentum goes into the excitation of fermionic edge modes.
Their velocity is reduced by the presence of interactions.

However, we show here data only for the $\nu=1$ Moore-Read state as smaller filling fractions do not converge sufficiently to be described by our effective theories at the system sizes we can reach with these exact methods.
We suspect that this is due to a larger correlation length in these systems\cite{estienne2015correlation}, which therefore requires higher order terms to be included in $H$ to provide an adequate description of the data at small system sizes.
This point is also true for the Laughlin state, whose agreement with our low energy theories also worsens as the filling fraction decreases.
We present an illustration of this point in Appendix \ref{small filling}.

\section{Conclusion}

We have found the behaviour of quantum Hall edge states in anharmonic traps and in the presence of short-range interactions and shown that  these theories are extremely accurate descriptions of the low energy structure of both the Laughlin and Moore-Read states.
These effective theories show that the addition of bulk translational symmetry causes surprising simplifications in the nonlinear Luttinger liquid expansion.
Furthermore, we find that these local descriptions of the edge behaviour are at odds with alternative proposals for the edge state behaviour relying on non-local models such as the quantum Benjamin-Ono equation\cite{wiegmann2012nonlinear}.

However, the present analysis does not cover the full behaviour of the resulting theories.
It would be extremely interesting to analyse the consequences of the remaining terms in greater details, considering for example the implications on the hydrodynamics of the systems, perhaps along similar lines to previous works\cite{bettelheim2006nonlinear, price2014fine, price2015quantum} which have, among other things, considered the potential for shockwaves along the edge.
Furthermore, the line of reasoning we use in this paper is readily applicable to any other quantum Hall wavefunction which can be expressed in terms of conformal blocks, which might indicate further interesting results in, for example, the Read-Rezayi states.
Finally, given that these results depend intimately on our conjecture of locality and that this is not fully understood, it would be worth exploring exactly how and why the Hamiltonian can be claimed to be local, something which we partially consider for the integer quantum Hall effect in a future publication\cite{bondesanFUTUREinteger}.

\section{Acknowledgements}

We are grateful to J\'er\^ome Dubail and Hans Hansson for enlightening discussions.
This work was supported by EPSRC grants EP/I031014/1 and EP/N01930X/1.
We would also like to acknowledge the DiagHam package, maintained by N. Regnault, which provided key assistance to this work.
Statement of compliance with EPSRC policy framework on research data: This publication is theoretical work that does not require supporting research data.

\bibliography{effectiveBib}

\appendix

\section{Haldane Pseudopotentials}
\label{Haldane pseudopotentials}

We provide a rapid review of the two-body problem and discuss the concept of Haldane pseudopotentials.
These pseudopotentials are a convenient way to think about potentials and generate parent Hamiltonians for the Laughlin state.
They are reviewed in \cite{girvin1999quantum} but we summarise the points pertinent to our discussion here.

The lowest Landau level of the single-particle quantum Hall effect is a set of degenerate states gapped by $\hbar\omega_c$ where $\omega_c$ is the cyclotron frequency.
As is customary\cite{girvin1984formalism} we use the Bargmann-Fock space of holomorphic polynomials as our Hilbert space with an inner product of the form
	\eq{\bbrakket{f}{g} = \int\frac{\de^2z}{\pi}e^{-\frac{|z|^2}{2\ell_B^2}} \overline{g(z)} f(z)}
where $\ell_B$ is the magnetic length and $z=x+iy$ is the particle position.
Note that we use curly kets to distinguish states in the physical space from those in the conformal field theory, which we introduce later.
Within this space the wavefunction $\bbrakket{z}{\psi_l} = z^l$ describes a particle with angular momentum $l$ about the origin.

We perturb the single-particle picture with a two-body interaction $V(|z_1-z_2|)\ll\hbar\omega_c$.
This perturbation is rotationally invariant and so is diagonal in a basis of states with well-defined relative and total angular momentum,
	\eq{\bbrakket{z_1,z_2}{m,M} = z_r^mZ^M}
where $z_r=z_1-z_2$ and $Z=\frac{z_1+z_2}{2}$ are the relative and centre-of-mass coordinates.
The interaction is also translationally invariant so the eigenvalues of $V$ are independent of $M$.

This leads to the concept of pseudopotentials.
We define the $m^\tr{th}$ pseudopotential of $V$ as
	\eq{v_m[V] = \frac{\bbra{m,M}V\kket{m,M}}{\bbrakket{m,M}{m,M}}.}
Therefore, given some set of functions $V_k(|z_r|)$ for which $v_m[V_k] = \delta_{m,k}$, we can represent any generic interaction potential in this Hilbert space as
	\eq{V(|z_r|) = \sum_{k=0}^\infty v_k[V] V_k(|z_r|).}
These $V_k(|z|)$ may then be expressed as the derivative of a delta function,
	\eq{V_k(|z_r|) = L_k\left(-\ell_B^2\nabla_r^2\right)\left[4\pi\ell_B^2\delta^{(2)}(z_r)\right]}
where $L_k$ is the $k^\tr{th}$ Laguerre polynomial.

\section{Smaller Fillings}
\label{small filling}

We present a series of plots to demonstrate the agreement between our effective Hamiltonians and numerics for a variety of filling fractions in the Laughlin state.
In each case we take only the least irrelevant term in the Hamiltonian, neglecting all other terms,
	\eq{H = g_{22}T_{22} + \order{R^{-5}}.}
Then, as in the main text, we fit the coupling coefficient, $g_{22}$, using the matrix element $\bra{0}a_2Ha_{-2}\ket{0}$ which, in theory, should depend only on $g_{22}$ to all orders.
Plotted are the subsequent agreements between this simple fit and the numerics in the case where the interactions we take in each case are the first contributing pseudopotentials for fillings $\nu=1/2$, $1/3$ and $1/4$, each at a system size of $N=8$ (which constitutes an approximate limit for our exact numerical methods in the $\nu=1/4$ state).
They show that the agreement becomes worse for smaller filling fractions, which also corresponds to larger correlation lengths and therefore one might expect to require a larger number of higher order terms to achieve an adequate agreement with the data.

\begin{figure}[ht!]
	\centering
	\includegraphics[width=0.92\linewidth]{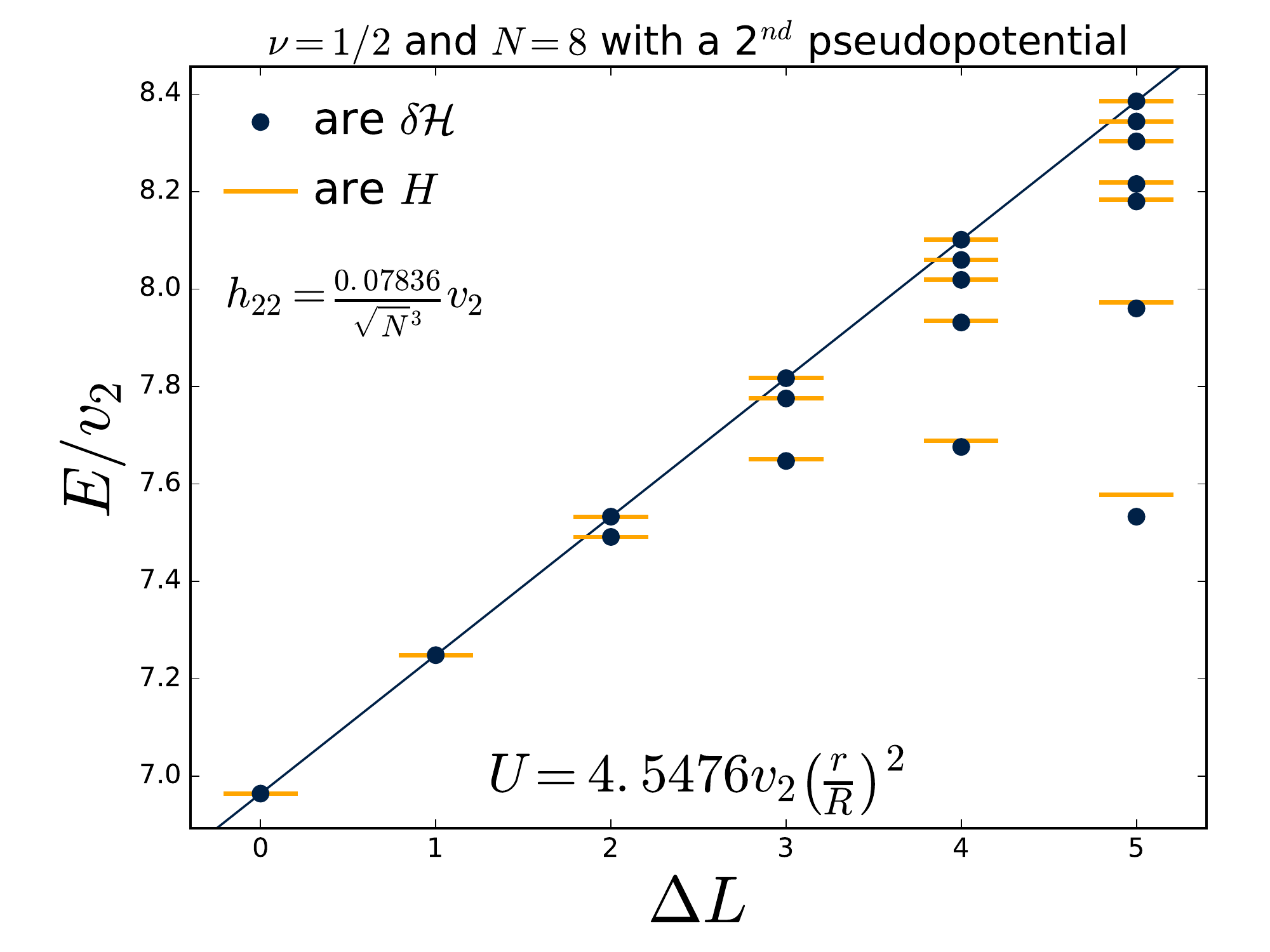}
	\includegraphics[width=0.92\linewidth]{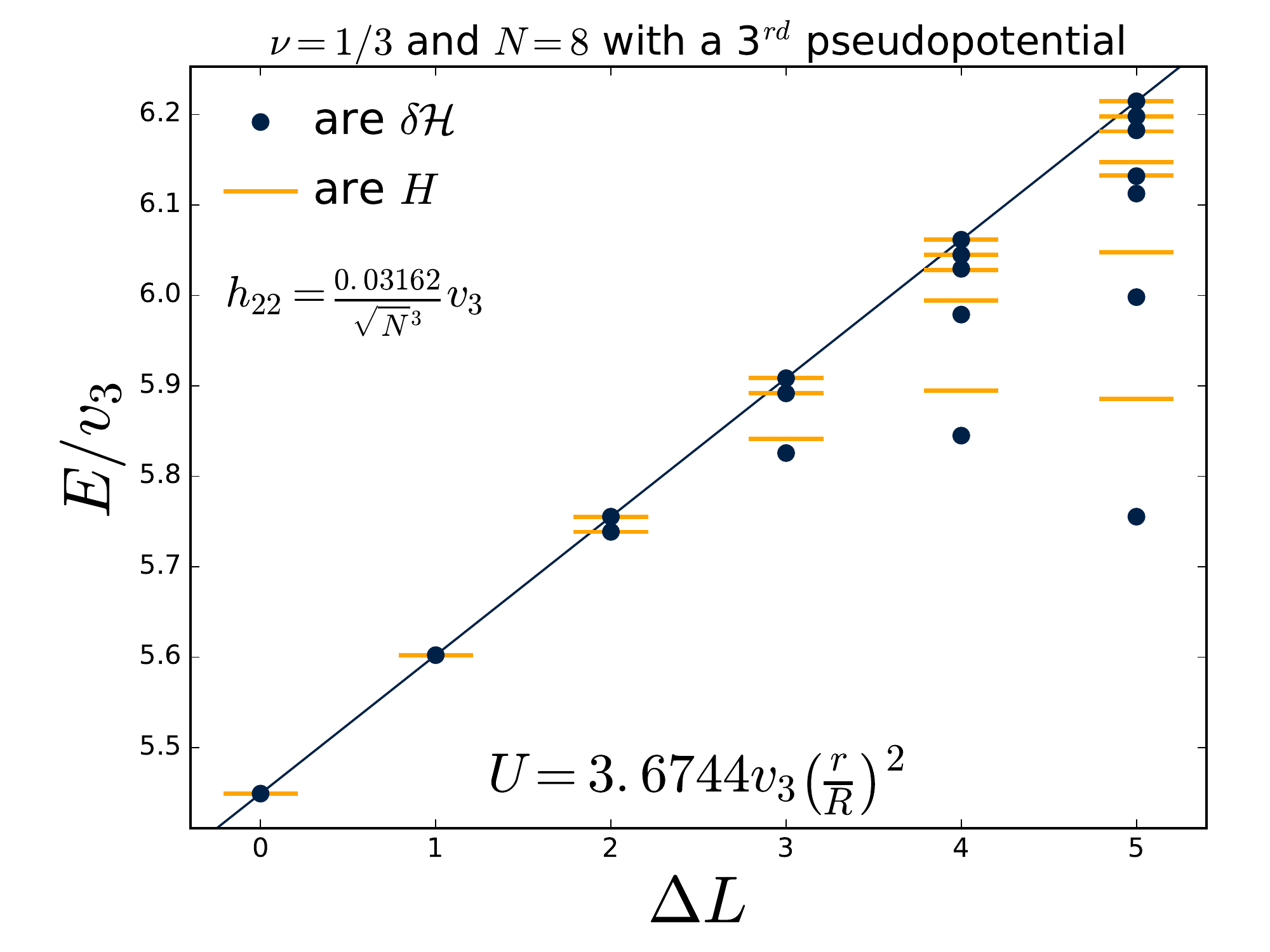}
	\includegraphics[width=0.92\linewidth]{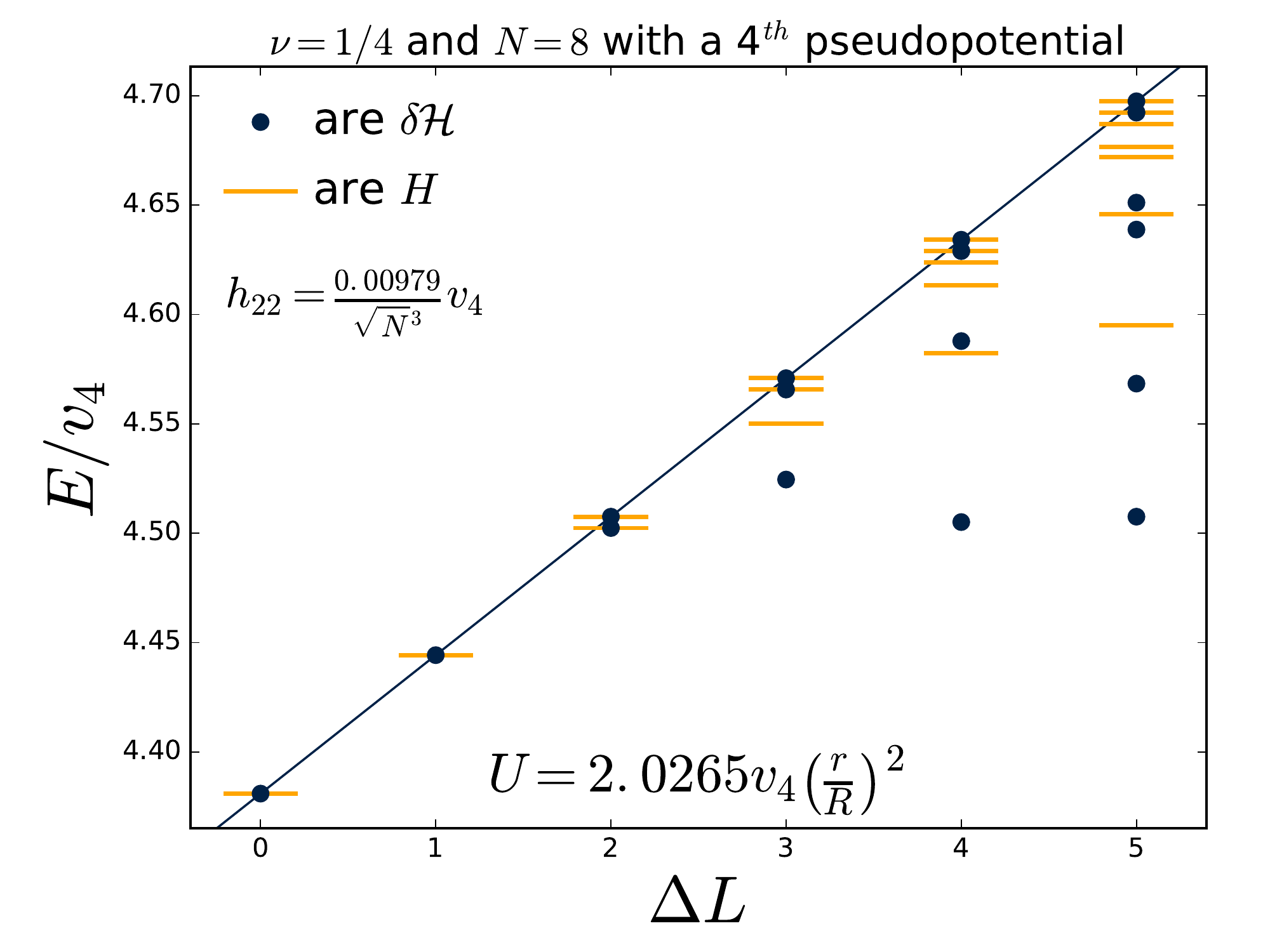}
	\caption{A trio of plots showing the spectrum for the quantum Hall edge when the bulk interactions are the first non-vanishing pseudopotentials at filling fractions $\nu=1/2$, $1/3$ and $1/4$, (i.e, at filling $1/m$ we add the pseudopotential $V_k$) and the system size is $N=8$.
	The effective Hamiltonian we fit is simply $H=g_{22}T_{22}$, with the coefficient fit using only the data at $\Delta L=2$.
	The blue points are the numerical data and the orange levels are the result of diagonalising the effective Hamiltonian.
	This data shows that the agreement becomes steadily worse as the filling fraction decreases, with the agreement very poor for $\nu=1/4$.
	To describe this case well one would require higher order terms in the effective Hamiltonian.}
	\label{small quarter}
\end{figure}

\section{Harmonic Interactions}
\label{Harmonic Interactions}

Consider the Harmonic interaction, which we take to have the form
	\eq{\delta\ham = V_0\sum_{i\neq j}\left|\frac{z_i-z_j}{2\ell_B}\right|^2.}
This interaction is clearly very non-local, coupling particles with larger separations more than those which are close.
In order to find the mapping of this operator onto the CFT we must first convert it into a differential operator using projection to the lowest Landau level $\bar z_i\to 2\ell_B^2\partial_i$.
The result of this is that
	\eq{\delta\ham = NV_0\sum_iz_i\partial_i-V_0\sum_iz_i\sum_j\partial_j + V_0N(N-1).}
Therefore, we simply need the mapping of these operators into the CFT.
These are all derived in the main text with
	\alg{\sum_iz_i & \to \frac{1}{\sqrt\beta}a_1 \\
		\sum_iz_i\partial_i & \to L_0 + \frac{\beta N(N-1)}{2} \\
		\sum_j\partial_j & \to L_{-1} + N\sqrt\beta a_{-1}}
in the sector with zero charge (i.e, $a_0=0$).
Collating these results we therefore find that
	\alg{H = NV_0&\lr{L_0 - a_1a_{-1} - \frac{1}{N\sqrt\beta}a_1L_{-1}} \nonumber\\
		 & + \frac{1}{2}V_0N(N-1)(N\beta+2).}
Note that this effective Hamiltonian is clearly non-local with the $a_1L_{-1}$ term being of the form
	\eq{a_1 L_{-1} = \frac{1}{2}\oint\cint{z}\oint\cint{w}zi\partial\varphi(z):\lr{i\partial\varphi(w)}^2:.}

\end{document}